\documentclass[12pt,twoside, a4paper]{article}
\def\pd{\partial}
\def\mc{\mathcal}
\def\ul{\underline}

\usepackage[dvips]{graphicx}
\usepackage{amssymb}
\usepackage[bbgreekl]{mathbbol}
\usepackage{amssymb,amsmath}
\usepackage{graphicx}
\usepackage{dsfont}
\usepackage{caption}
\usepackage{subcaption}
\usepackage{mathtools}
\usepackage{verbatim}
\usepackage{graphicx}
\usepackage{multirow}
\usepackage[outline]{contour}
\usepackage{xcolor,colortbl}
\input{epsf.sty}  
\begin{document}
\begin{center}
\Large{\textbf{New supersymmetric $AdS_6$ black holes from matter-coupled $F(4)$ gauged supergravity}}
\end{center}
\vspace{1 cm}
\begin{center}
\large{\textbf{Parinya Karndumri}}
\end{center}
\begin{center}
String Theory and Supergravity Group, Department
of Physics, Faculty of Science, Chulalongkorn University, 254 Phayathai Road, Pathumwan, Bangkok 10330, Thailand \\
E-mail: parinya.ka@hotmail.com \vspace{1 cm}
\end{center}
\begin{abstract}
We study supersymmetric $AdS_6$ black holes from matter-coupled $N=(1,1)$ gauged supergravity coupled to three vector multiplets and $SO(3)\times SO(3)$ gauge group. This gauged supergravity admits two supersymmetric $AdS_6$ vacua preserving all supersymmetries with $SO(3)\times SO(3)$ and $SO(3)_{\textrm{diag}}$ symmetries. By considering a truncation to $SO(2)_{\textrm{diag}}\subset SO(3)_{\textrm{diag}}$ invariant sector, we find a number of new supersymmetric $AdS_2\times \mc{M}_4$ solutions by performing a topological twist along $\mc{M}_4$. For $\mc{M}_4$ being a product of two Riemann surfaces $\Sigma\times \widetilde{\Sigma}$ and a Kahler four-cycle, the twist is implemented by $SO(2)_{\textrm{diag}}$ gauge field while, for $\mc{M}_4$ given by a Cayley four-cycle, the twist is performed by turning on $SO(3)_{\textrm{diag}}$ gauge fields. We give numerical black hole solutions interpolating between $AdS_2\times \mc{M}_4$ near horizon geometries and asymptotically locally $AdS_6$ vacua. Among these solutions, there are solutions interpolating between both of the supersymmetric $AdS_6$ vacua and near horizon geometries. The solutions can also be interpreted as holographic RG flows from five-dimensional SCFTs to superconformal quantum mechanics.   
\end{abstract}
\newpage
\tableofcontents
\newpage
\section{Introduction}
The study of supersymmetric asymptotically AdS black holes in various dimensions has attracted much attention since the understanding of the microscopic origin of the Bekenstein-Hawking entropy has been acquired by the computation of topologically twisted indices in the dual field theories \cite{Twisted_index1,Twisted_index2,Twisted_index3}, see also \cite{BH_counting1}-\cite{BH_counting13} for results on different types of black holes in various dimensions. The procedure has also been generalized to black strings in five dimensions in \cite{black_string_entropy}. In general, the black hole solutions interpolate between an asymptotically $AdS_{d+1}$ space and a near horizon geometry $AdS_2\times \mc{M}_{d-1}$ with $\mc{M}_{d-1}$ being the event horizon of the black holes. According to the AdS/CFT correspondence \cite{maldacena}, these solutions also describe holographic RG flows across dimensions from $d$-dimensional SCFTs dual to $AdS_{d+1}$ in the UV to superconformal quantum mechanics dual to the $AdS_2$ space in the IR via twisted compactifications on $\mc{M}_{d-1}$. 
\\
\indent In six dimensions, only the half-maximal non-chiral $N=(1,1)$ or $F(4)$ gauged supergravity constructed in \cite{F4_Romans} admits supersymmetric $AdS_6$ vacua. The matter-coupled $F(4)$ gauged supergravity has been constructed in \cite{F4SUGRA1,F4SUGRA2}, and general conditions on the existence of supersymmetric $AdS_6$ vacua have been given in \cite{AdS6_3}. The corresponding $AdS_6$ black hole solutions with the near horizon geometry of the form $AdS_2\times \mc{M}_4$ from pure $F(4)$ gauged supergravity has been considered in \cite{Naka} and more recently in \cite{AdS6_BH_Minwoo1} and \cite{flow_across_bobev}. More general solutions in the matter-coupled $F(4)$ gauged supergravity have appeared later in \cite{AdS6_BH_Zaffaroni} and \cite{AdS6_BH_Minwoo}. In \cite{AdS6_BH_Zaffaroni}, the $F(4)$ gauged supergravity is coupled to one vector multiplet leading to $SO(3)\times U(1)$ gauge group with all the fields being uncharged under the $U(1)$ factor. A more general $F(4)$ gauged supergravity coupled to three vector multiplets has been considered in \cite{AdS6_BH_Minwoo}. The resulting gauged supergravity has $SO(3)\times SO(3)$ gauge group. However, the black hole solutions with $AdS_2\times H^2\times H^2$ and $AdS_2\times \mc{M}_4$ for $\mc{M}_4$ being a Kahler four-cycle given in \cite{AdS6_BH_Minwoo} have been found by considering a truncation to $SO(2)\times SO(2)$ invariant sector. Within this sector, the effect of the second $SO(3)$ factor in the gauge group is invisible. In particular, the gauge coupling constant for this factor does not appear at all in the BPS equations.       
\\
\indent In this work, we look for more interesting supersymmetric $AdS_6$ black hole solutions within this matter-coupled $F(4)$ gauged supergravity with three vector multiplets and $SO(3)\times SO(3)$ gauge group. This gauged supergravity has been first studied in \cite{F4_flow}, and further in \cite{5DSYM_from_F4}, in which two $N=(1,1)$ supersymmetric $AdS_6$ vacua with $SO(3)\times SO(3)$ and $SO(3)_{\textrm{diag}}$ symmetries have been found together with holographic RG flows between $AdS_6$ vacua and RG flows to non-conformal phases of the dual five-dimensional SCFTs. In particular, supersymmetric $AdS_4\times \Sigma$ with $\Sigma=S^2, H^2$ and $AdS_3\times \mc{M}_3$ with $\mc{M}_3=S^3, H^3$ solutions have been studied in \cite{6D_twist}. These solutions correspond to near horizon geometries of black strings and black two-branes in asymptotically $AdS_6$ spaces. 
\\
\indent In contrast to the previous study of $SO(2)\times SO(2)$ invariant sector, we will consider a smaller residual symmetry $SO(2)_{\textrm{diag}}\subset SO(2)\times SO(2)$ and look for black holes with near horizon geometries of the form $AdS_2\times \mc{M}_4$ for $\mc{M}_4$ being a product of two Riemann surfaces or a Kahler four-cycle. We will see that, for negatively curved $\mc{M}_4$ spaces, there exist a number of supersymmetric black hole solutions in asymptotically $AdS_6$ spaces described by the two aforementioned $AdS_6$ vacua. In addition, we will also consider the black hole solutions with $\mc{M}_4$ given by a Cayley four-cycle by performing a topological twist using $SO(3)_{\textrm{diag}}$ gauge fields. Although this has already been studied in \cite{AdS6_BH_Minwoo}, we indeed find new supersymmetric $AdS_6$ black hole solutions interpolating between the near horizon geometry and the two supersymmetric $AdS_6$ vacua.  
\\
\indent The paper is organized as follows. We give a brief review of
the matter-coupled $F(4)$ gauged supergravity in section \ref{6D_SO4gaugedN2}. We then consider the $F(4)$ gauged supergravity coupled to three vector multiplets with $SO(3)\times SO(3)$ gauge group and review two known supersymmetric $AdS_6$ vacua. By truncating to $SO(2)_{\textrm{diag}}$ invariant sector, we study $AdS_2\times \Sigma\times\widetilde{\Sigma}$ and find a new class of $AdS_2\times H^2\times H^2$ solutions in section \ref{AdS2_Sigma_Sigma}. A number of numerical black hole solutions are also given. We subsequently extend the anlysis to the case of $AdS_2\times \mc{M}_4$ solutions with $\mc{M}_4$ given by a Kahler four-cycle in section \ref{AdS2_M4}. We also consider the case of $\mc{M}_4$ being a Cayley four-cycle by performing a topological twist using $SO(3)_{\textrm{diag}}$ gauge fields. In this case, we find a new $AdS_6$ black hole solution in addition to the solution found previously in \cite{AdS6_BH_Minwoo}. We finally give some conclusions and comments in section \ref{conclusion}. In the appendix, all the bosonic field equations of the matter-coupled $F(4)$ gauged supergravity are recorded. 
\section{Matter coupled $N=(1,1)$ gauged supergravity in six dimensions}\label{6D_SO4gaugedN2}
In this section, we review the structure of matter-coupled $F(4)$ gauged supergravity in six dimensions obtained by gauging the half-maximal $N=(1,1)$ supergravity coupled to vector multiplets. We mostly follow the conventions of the original construction in
\cite{F4SUGRA1, F4SUGRA2} but with the metric signature $(-+++++)$. We firstly review the general matter-coupled $F(4)$ gauged supergravity coupled to an arbitrary number $n$ of vector multiplets and finally consider the case of $n=3$ and $SO(3)\times SO(3)$ gauge group. 

\subsection{General structure of matter-coupled $F(4)$ gauged supergravity}
The $N=(1,1)$ supergravity multiplet in six dimensions consists of the following component fields 
\begin{displaymath}
\left(e^a_\mu,\psi^A_\mu, A^\alpha_\mu, B_{\mu\nu}, \chi^A,
\sigma\right).
\end{displaymath}
The bosonic fields are the graviton $e^a_\mu$, a two-form field $B_{\mu\nu}$, four vector fields $A^\alpha_\mu,\, \alpha=0,1,2,3$, and the dilaton $\sigma$. Space-time and tangent space or flat indices are
denoted respectively by $\mu,\nu=0,\ldots ,5$ and $a,b=0,\ldots, 5$. The fermionic fields are given by two gravitini $\psi^A_\mu$ and two spin-$\frac{1}{2}$ fields $\chi^A$ with indices $A,B,\ldots =1,2$ denoting the fundamental representation of $SU(2)_R\sim USp(2)_R\sim SO(3)_R$ R-symmetry. Following \cite{F4SUGRA1} and \cite{F4SUGRA2}, we also introduce the $SU(2)_R$ adjoint indices $r,s,\ldots =1,2,3$ according to the split of indices $\alpha=(0,r)$.
\\
\indent The vector multiplets are described by the field content
\begin{displaymath}
(A_\mu,\lambda_A,\phi^\alpha)^I, \qquad I=1,2,\ldots, n
\end{displaymath}
consisting of $n$ vectors $A^I_\mu$, $2n$ gaugini $\lambda^I_A$, and $4n$ scalars $\phi^{\alpha I}$ parametrizing $SO(4,n)/SO(4)\times SO(n)$ coset manifold. All spinor fields $\chi^A$, $\psi^A_\mu$ and
$\lambda_A$ as well as the supersymmetry parameter $\epsilon^A$ are
eight-component pseudo-Majorana spinors. In addition, the $4+n$ vector fields from both the gravity and vector multiplets will be collectively denoted by $A^\Lambda=(A^\alpha,A^I)$.
\\
\indent Including the dilaton, there are $4n+1$ scalar fields described by
$\mathbb{R}^+\times SO(4,n)/SO(4)\times SO(n)$ coset with $\mathbb{R}^+$ corresponding to the dilaton. The $4n$ vector multiplet scalars can be parametrized by a coset
representative ${L^\Lambda}_{\ul{\Sigma}}$ transforming under the global $SO(4,n)$ and local $SO(4)\times SO(n)$ symmetries by left and right multiplications respectively with indices $\Lambda,\ul{\Sigma}=0,\ldots , n+3$. We can also split the index $\ul{\Sigma}$ transforming under the local $SO(4)\times
SO(n)$ as $\ul{\Sigma}=(\alpha,I)=(0,r,I)$. Accordingly, the coset representative can be written as
\begin{equation}
{L^\Lambda}_{\ul{\Sigma}}=({L^\Lambda}_\alpha,{L^\Lambda}_I).
\end{equation}
The inverse of ${L^\Lambda}_{\ul{\Sigma}}$ will be denoted by ${(L^{-1})^{\ul{\Lambda}}}_\Sigma=({(L^{-1})^{\alpha}}_\Sigma,{(L^{-1})^{I}}_\Sigma)$. $SO(4,n)$ indices will be raised and lowered by the invariant tensor
\begin{equation}
\eta_{\Lambda\Sigma}=\eta^{\Lambda\Sigma}=(\delta_{\alpha\beta},-\delta_{IJ}).
\end{equation}
\indent We are interested in gauging a compact subgroup of $SO(4,n)$ of the form $G=SO(3)\times G_c$ in which the $SO(3)$ factor is identified with the R-symmetry $SO(3)_R\sim SU(2)_R$. This $SO(3)$ is gauged by three vector fields $A^r$ within the gravity multiplet. $G_c$ is a compact subgroup of $SO(n)$ gauged by the vector fields in vector multiplets with $\textrm{dim}\, (G_c)\leq n$. The structure constants of the gauge algebra ${f^\Lambda}_{\Sigma\Gamma}$ appearing in the Lie algebra of the gauge generators $T_\Lambda$ as  
\begin{equation}
\left[T_\Lambda,T_\Sigma\right]={f^\Gamma}_{\Lambda\Sigma}T_\Gamma\qquad \textrm{with}\qquad f_{[\Lambda\Sigma\Gamma]}=0   
\end{equation}   
can be chosen to be $(\epsilon_{rst},C_{IJK})$ with $C_{IJK}$ being the structure constants of $G_c$.        
\\
\indent In addition to the gauging of $G\subset SO(4,n)$, there is also a massive deformation of the two-form field needed for the existence of $AdS_6$ vacua. With both of these deformations taken into account, the Lagrangian for $N=(1,1)$ gauged supergravity can be written as
\begin{eqnarray}
e^{-1}\mathcal{L}&=&\frac{1}{4}R-e\pd_\mu \sigma\pd^\mu \sigma
-\frac{1}{4}P^{I\alpha}_\mu P^{\mu}_{I\alpha}-\frac{1}{8}e^{-2\sigma}\mc{N}_{\Lambda\Sigma}\widehat{F}^\Lambda_{\mu\nu}
\widehat{F}^{\Sigma\mu\nu}-\frac{3}{64}e^{4\sigma}H_{\mu\nu\rho}H^{\mu\nu\rho}\nonumber \\
& &-V-\frac{1}{64}e^{-1}\epsilon^{\mu\nu\rho\sigma\lambda\tau}B_{\mu\nu}\left(\eta_{\Lambda\Sigma}\widehat{F}^\Lambda_{\rho\sigma}
\widehat{F}^\Sigma_{\lambda\tau}+mB_{\rho\sigma}\widehat{F}^\Lambda_{\lambda\tau}\delta^{\Lambda 0}+\frac{1}{3}m^2B_{\rho\sigma}B_{\lambda\tau}\right)\nonumber \\
\label{Lar}
\end{eqnarray}
with $e=\sqrt{-g}$. Various field strength tensors are defined by
\begin{eqnarray}
\widehat{F}^\Lambda=F^\Lambda-m\delta^{\Lambda 0}B, \qquad F^\Lambda=dA^\Lambda+\frac{1}{2}{f^\Lambda}_{\Sigma\Gamma} A^\Sigma\wedge A^\Gamma,\qquad H=dB\, .
\end{eqnarray}
It is also useful to note the convention on components of form fields used in \cite{F4SUGRA1, F4SUGRA2}
\begin{equation}
F^\Lambda=F^\Lambda_{\mu\nu} dx^\mu \wedge dx^\nu\qquad \textrm{and}\qquad H=H_{\mu\nu\rho}dx^\mu \wedge dx^\nu\wedge dx^\rho\, .
\end{equation}
In particular, these lead to for example $F^\Lambda_{\mu\nu}=\frac{1}{2}\left(\pd_\mu A_\nu-\pd_\nu A_\mu +\frac{1}{2}{f^\Lambda}_{\Sigma\Gamma}A^\Sigma_\mu A^\Gamma_\nu\right)$.  
\\
\indent The scalar kinetic term is written in terms of the vielbein on $SO(4,n)/SO(4)\times SO(n)$ denoted by $P^{I\alpha}_\mu=P^{I\alpha}_x\pd_\mu\phi^x$, $x=1,\ldots, 4n$. This vielbein together with the $SO(4)\times SO(n)$ composite connections $(\Omega^{rs},\Omega^{r0},\Omega^{IJ})$ are encoded in the left-invariant 1-form
\begin{equation}
{\Omega^{\ul{\Lambda}}}_{\ul{\Sigma}}=
{(L^{-1})^{\ul{\Lambda}}}_{\Pi}\nabla {L^\Pi}_{\ul{\Sigma}}
\qquad \textrm{with}\qquad \nabla
{L^\Lambda}_{\ul{\Sigma}}={dL^\Lambda}_{\ul{\Sigma}}
-f^{\phantom{\Gamma}\Lambda}_{\Gamma\phantom{\Lambda}\Pi}A^\Gamma
{L^\Pi}_{\ul{\Sigma}}\, .
\end{equation}
This leads to the vielbein with the following identification  
\begin{equation}
P^I_{\phantom{s}\alpha}=(P^I_{\phantom{a}0},P^I_{\phantom{a}r})=(\Omega^I_{\phantom{a}0},\Omega^I_{\phantom{a}r}).
\end{equation}
The symmetric scalar matrix $\mc{N}_{\Lambda\Sigma}$ appearing in the kinetic term of the vector fields is defined by
\begin{eqnarray}
\mc{N}_{\Lambda\Sigma}=L_{\Lambda\alpha}{(L^{-1})^\alpha}_\Sigma-L_{\Lambda I}{(L^{-1})^I}_\Sigma=(\eta L L^T\eta)_{\Lambda\Sigma}\, .
\end{eqnarray}
\indent As usual, gaugings lead to the scalar potential and modified supersymmetry transformations of fermions by a number of fermion-shift matrices. The explicit form of the scalar potential reads
\begin{eqnarray}
V&=&-e^{2\sigma}\left[\frac{1}{36}A^2+\frac{1}{4}B^iB_i+\frac{1}{4}\left(C^I_{\phantom{s}t}C_{It}+4D^I_{\phantom{s}t}D_{It}\right)\right]
+m^2e^{-6\sigma}\mc{N}_{00}\nonumber \\
& &-me^{-2\sigma}\left[\frac{2}{3}AL_{00}-2B^iL_{0i}\right]
\end{eqnarray}
with $\mc{N}_{00}$ being the $00$ component of $\mc{N}_{\Lambda\Sigma}$. Supersymmetry transformation rules for all the fermionic fields are given by
\begin{eqnarray}
\delta\psi_{\mu
A}&=&D_\mu\epsilon_A-\frac{1}{24}\left(Ae^\sigma+6me^{-3\sigma}(L^{-1})_{00}\right)\epsilon_{AB}\gamma_\mu\epsilon^B\nonumber
\\
& &-\frac{1}{8}
\left(B_te^\sigma-2me^{-3\sigma}(L^{-1})_{t0}\right)\gamma^7\sigma^t_{AB}\gamma_\mu\epsilon^B\nonumber \\
&
&+\frac{i}{16}e^{-\sigma}\left[\epsilon_{AB}(L^{-1})_{0\Lambda}\gamma_7+\sigma^r_{AB}(L^{-1})_{r\Lambda}\right]
F^\Lambda_{\nu\lambda}(\gamma_\mu^{\phantom{s}\nu\lambda}
-6\delta^\nu_\mu\gamma^\lambda)\epsilon^B\nonumber \\
& &+\frac{i}{32}e^{2\sigma}H_{\nu\lambda\rho}\gamma_7({\gamma_\mu}^{\nu\lambda\rho}-3\delta_\mu^\nu\gamma^{\lambda\rho})\epsilon_A,\label{delta_psi}\\
\delta\chi_A&=&\frac{1}{2}\gamma^\mu\pd_\mu\sigma\epsilon_{AB}\epsilon^B+\frac{1}{24}
\left[Ae^\sigma-18me^{-3\sigma}(L^{-1})_{00}\right]\epsilon_{AB}\epsilon^B\nonumber
\\
& &-\frac{1}{8}
\left[B_te^\sigma+6me^{-3\sigma}(L^{-1})_{t0}\right]\gamma^7\sigma^t_{AB}\epsilon^B\nonumber
\\
& &-\frac{i}{16}e^{-\sigma}\left[\sigma^r_{AB}(L^{-1})_{r\Lambda}-\epsilon_{AB}(L^{-1})_{0\Lambda}\gamma_7\right]F^\Lambda_{\mu\nu}\gamma^{\mu\nu}\epsilon^B\nonumber \\
& &-\frac{i}{32}e^{2\sigma}H_{\nu\lambda\rho}\gamma_7\gamma^{\nu\lambda\rho}\epsilon_A,
\label{delta_chi}\\
\delta
\lambda^{I}_A&=&P^I_{ri}\gamma^\mu\pd_\mu\phi^i\sigma^{r}_{\phantom{s}AB}\epsilon^B+P^I_{0i}
\gamma^7\gamma^\mu\pd_\mu\phi^i\epsilon_{AB}\epsilon^B-\left(2i\gamma^7D^I_{\phantom{s}t}+C^I_{\phantom{s}t}\right)
e^\sigma\sigma^t_{AB}\epsilon^B \nonumber
\\
& &+2me^{-3\sigma}(L^{-1})^I_{\phantom{ss}0}
\gamma^7\epsilon_{AB}\epsilon^B-\frac{i}{2}e^{-\sigma}(L^{-1})^I_{\phantom{s}\Lambda}F^\Lambda_{\mu\nu}
\gamma^{\mu\nu}\epsilon_{A}\label{delta_lambda}
\end{eqnarray}
where $\sigma^{tC}_{\phantom{sd}B}$ are usual Pauli matrices, and
$\epsilon_{AB}=-\epsilon_{BA}$. $A,B$ indices can be raised and lowered by $\epsilon^{AB}$ and $\epsilon_{AB}$ with the convention $T^A=\epsilon^{AB}T_B$ and $T_A=T^B\epsilon_{BA}$. The covariant derivative of $\epsilon_A$ is
given by
\begin{equation}
D_\mu \epsilon_A=\pd_\mu
\epsilon_A+\frac{1}{4}\omega_\mu^{ab}\gamma_{ab}\epsilon_A+\frac{i}{2}\sigma^r_{AB}
\left[\frac{1}{2}\epsilon^{rst}\Omega_{\mu st}-i\gamma_7
\Omega_{\mu r0}\right]\epsilon^B\, .
\end{equation}
\indent Various components of the fermion-shift matrices are defined as follows
\begin{eqnarray}
A&=&\epsilon^{rst}K_{rst},\qquad B^i=\epsilon^{ijk}K_{jk0},\\
C^{\phantom{ts}t}_I&=&\epsilon^{trs}K_{rIs},\qquad D_{It}=K_{0It}
\end{eqnarray}
where
\begin{eqnarray}
K_{rst}&=&g_1\epsilon_{lmn}L^l_{\phantom{r}r}(L^{-1})_s^{\phantom{s}m}L_{\phantom{s}t}^n+
g_2C_{IJK}L^I_{\phantom{r}r}(L^{-1})_s^{\phantom{s}J}L_{\phantom{s}t}^K,\nonumber
\\
K_{rs0}&=&g_1\epsilon_{lmn}L^l_{\phantom{r}r}(L^{-1})_s^{\phantom{s}m}L_{\phantom{s}0}^n+
g_2C_{IJK}L^I_{\phantom{r}r}(L^{-1})_s^{\phantom{s}J}L_{\phantom{s}0}^K,\nonumber
\\
K_{rIt}&=&g_1\epsilon_{lmn}L^l_{\phantom{r}r}(L^{-1})_I^{\phantom{s}m}L_{\phantom{s}t}^n+
g_2C_{IJK}L^I_{\phantom{r}r}(L^{-1})_I^{\phantom{s}J}L_{\phantom{s}t}^K,\nonumber
\\
K_{0It}&=&g_1\epsilon_{lmn}L^l_{\phantom{r}0}(L^{-1})_I^{\phantom{s}m}L_{\phantom{s}t}^n+
g_2C_{IJK}L^I_{\phantom{r}0}(L^{-1})_I^{\phantom{s}J}L_{\phantom{s}t}^K\,
.
\end{eqnarray}
\indent Finally, we note the convention on space-time
gamma matrices $\gamma^a$ which is slightly different from those of \cite{F4SUGRA1, F4SUGRA2}. $\gamma^a$ satisfy the Clifford algebra
\begin{equation}
\{\gamma^a,\gamma^b\}=2\eta^{ab},\qquad
\eta^{ab}=\textrm{diag}(-1,1,1,1,1,1),
\end{equation}
and the chirality matrix is defined by $\gamma_7=i\gamma^0\gamma^1\gamma^2\gamma^3\gamma^4\gamma^5$ with
$\gamma_7^2=-\mathbf{1}$. 

\subsection{Matter-coupled $F(4)$ gauged supergravity with $SO(3)\times SO(3)$ gauge group and supersymmetric $AdS_6$ vacua}
We now consider a specific case of $n=3$ vector multiplets and $SO(3)\times SO(3)\sim SU(2)\times SU(2)$ gauge group. As previously mentioned, the first $SO(3)$ is the R-symmetry gauged by $A^r_\mu$, and the second one gauged by $A^I_\mu$, $I=1,2,3$, from the three vector multiplets. With $C_{IJK}=\epsilon_{IJK}$, the structure constants of the full gauge group are then given by 
\begin{equation}
{f^\Lambda}_{\Pi\Sigma}=(\epsilon_{rst},\epsilon_{IJK}). 
\end{equation}
The explicit parametrization of the scalar fields in $SO(4,3)/SO(4)\times SO(3)$ coset can be obtained as in \cite{F4_flow} by introducing the $7\times 7$ matrices
\begin{equation}
(e^{\Lambda \Sigma})_{\Gamma \Pi}=\delta^\Lambda_{
\Gamma}\delta^\Sigma_{\Pi},\qquad \Lambda, \Sigma,\Gamma,
\Pi=0,\ldots ,6\, .
\end{equation}
The compact $SO(4)\times SO(3)$ generators are given by
\begin{eqnarray}
SO(4):\qquad
J^{\alpha\beta}&=&e^{\beta,\alpha}-e^{\alpha,\beta},\qquad \alpha,\beta=0,1,2,3,\nonumber \\
SO(3):\qquad \tilde{J}^{IJ}&=&e^{J+3,I+3}-e^{I+3,J+3},\qquad I,J=1,2,3
\end{eqnarray}
while non-compact generators can be identified as
\begin{eqnarray}
Y_{\alpha I}=e^{\alpha,I+3}+e^{I+3,\alpha}\, .
\end{eqnarray}
The structure constants given above imply that the $SO(3)\times SO(3)$ gauge generators are given respectively by $J^{rs}$ and $\tilde{J}^{IJ}$. 
\\
\indent This gauged supergravity has been originally studied in \cite{F4_flow}, and two supersymmetric $N=(1,1)$ $AdS_6$ vacua with $SO(3)\times SO(3)$ and $SO(3)_{\textrm{diag}}$ symmetries have been identified. For convenience, we will also present these two vacua here. With the $SO(3)_{\textrm{diag}}$ generated by $J^{rs}+\tilde{J}^{rs}$, the only one singlet scalar from $SO(4,3)/SO(4)\times SO(3)$ coset corresponds to the non-compact generator $Y_{11}+Y_{22}+Y_{33}$. The coset representative can be written as 
\begin{equation}
L=e^{\phi(Y_{11}+Y_{22}+Y_{33})}\, .\label{SO3diag_L}
\end{equation}
The resulting scalar potential reads
\begin{eqnarray}
V&=&\frac{1}{16}e^{2\sigma}\left[(g_1^2+g_2^2)[\cosh 6\phi-9\cosh 2\phi]+8(g_2^2-g_1^2)+8g_1g_2
\sinh^32\phi\right]\nonumber \\
& &+e^{-6\sigma}m^2-4e^{-2\sigma}m(g_1\cosh^3\phi-g_2\sinh^3\phi)
\end{eqnarray}
which admits two supersymmetric $AdS_6$ critical points given by
\begin{equation}
\phi=0,\qquad \sigma=\frac{1}{4}\ln
\left[\frac{3m}{g_1}\right],\qquad
V_0=-20m^2\left(\frac{g_1}{3m}\right)^{\frac{3}{2}}\label{SO4_AdS6}
\end{equation}
and
\begin{eqnarray}
\phi&=&\frac{1}{2}\ln \left[\frac{g_1+g_2}{g_2-g_1}\right],\qquad
\sigma=\frac{1}{4}\ln
\left[\frac{3m\sqrt{g_2^2-g_1^2}}{g_1g_2}\right],\nonumber \\
V_0&=&-20m^2\left[\frac{g_1g_2}{3m\sqrt{g_2^2-g_1^2}}\right]^{\frac{3}{2}}\,
.\label{SO3_AdS6}
\end{eqnarray}
The first critical point is $SO(4)\sim SO(3)\times SO(3)$ invariant while the second one preserves only $SO(3)_{\textrm{diag}}\subset SO(3)\times SO(3)$. For later convenience, we will refer to these vacua as $AdS_6$ critical point $i$ and $ii$, respectively. We can also set $g_1=3m$ to have vanishing dilaton at critical point $i$.

\section{Supersymmetric $AdS_6$ black holes with $\Sigma\times \widetilde{\Sigma}$ horizons}\label{AdS2_Sigma_Sigma}
We first consider black hole solutions with the near horizon geometry of the form $AdS_2\times \Sigma\times \widetilde{\Sigma}$ for $\Sigma$ and $\widetilde{\Sigma}$ being Riemann surfaces. The metric ansatz takes the
form of
\begin{equation}
ds^2=-e^{2f(r)}dt^2+dr^2+e^{2h(r)}(d\theta^2+F_\kappa(\theta)^2 d\phi^2)+e^{2\tilde{h}(r)}(d\tilde{\theta}^2+\tilde{F}_{\tilde{\kappa}}(\tilde{\theta})^2 d\tilde{\phi}^2)
\end{equation}
with the function $F_\kappa(\theta)$ given by
\begin{equation}
F_\kappa(\theta)=\begin{cases}
  \sin\theta,  & \kappa=1\phantom{-}\quad \textrm{for}\quad \Sigma^2=S^2 \\
  \theta,  & \kappa=0\phantom{-}\quad \textrm{for}\quad \Sigma^2=T^2\\
  \sinh\theta,  & \kappa=-1\quad \textrm{for}\quad \Sigma^2=H^2
\end{cases}\label{F_def}
\end{equation}
and similarly for $\tilde{F}_{\tilde{\kappa}}(\tilde{\theta})$. 
\\
\indent With the following choice of vielbein 
\begin{eqnarray}
& &e^{\hat{t}}=e^fdt,\qquad e^{\hat{r}}=dr,\qquad e^{\hat{\theta}}=e^hd\theta,\nonumber \\
& &e^{\hat{\phi}}=e^hF_\kappa(\theta) d\phi,\qquad e^{\hat{\tilde{\theta}}}=e^{\tilde{h}}d\tilde{\theta},\qquad e^{\hat{\tilde{\phi}}}=e^{\tilde{h}}\tilde{F}_{\tilde{\kappa}}(\tilde{\theta})d\tilde{\phi},
\end{eqnarray}
non-vanishing components of the spin connection for the above metric are
given explicitly by
\begin{eqnarray}
& &{\omega^{\hat{t}}}_{\hat{r}} =f'e^{\hat{t}},\qquad
{\omega^{\hat{\theta}}}_{\hat{r}}=h'e^{\hat{\theta}},\qquad {\omega^{\hat{\phi}}}_{\hat{r}}=h'e^{\hat{\phi}},\qquad {\omega^{\hat{\tilde{\theta}}}}_{\hat{r}}=\tilde{h}'e^{\hat{\tilde{\theta}}} \nonumber \\
& &{\omega^{\hat{\tilde{\phi}}}}_{\hat{r}}=\tilde{h}'e^{\hat{\tilde{\phi}}},\qquad {\omega^{\hat{\phi}}}_{\hat{\theta}}=\frac{F'_\kappa(\theta)}{F_\kappa(\theta)}e^{-h}e^{\hat{\phi}},\qquad
{\omega^{\hat{\tilde{\phi}}}}_{\hat{\tilde{\theta}}}=\frac{\tilde{F}'_{\tilde{\kappa}}(\tilde{\theta})}{\tilde{F}_{\tilde{\kappa}}(\tilde{\theta})}e^{-\tilde{h}}e^{\hat{\tilde{\phi}}}
\end{eqnarray}
with hatted indices being flat indices. Throughout the paper, we will use $'$ to denote $r$-derivatives except for $F'_\kappa(\theta)=\frac{dF_\kappa(\theta)}{d\theta}$ and $\tilde{F}'_{\tilde{\kappa}}(\tilde{\theta})=\frac{d\tilde{F}_{\tilde{\kappa}}(\tilde{\theta})}{d\tilde{\theta}}$. We also note some useful relations 
\begin{equation}
F''_\kappa(\theta)=-\kappa F_\kappa(\theta)\qquad \textrm{and}\qquad 1-{F'_\kappa(\theta)}^2=\kappa F_\kappa(\theta)^2
\end{equation}
which also hold for $\tilde{F}_{\tilde{\kappa}}(\tilde{\theta})$. 
\\
\indent In this section, we will consider a truncation to $SO(2)_{\textrm{diag}}$ invariant sector. There are four singlet scalars which, for $SO(2)_{\textrm{diag}}$ generated by $J^{12}+\tilde{J}^{12}$, can be described by the coset representative
\begin{equation}
L=e^{\phi_0 Y_{03}}e^{\phi_1 (Y_{11}+Y_{22})}e^{\phi_2 Y_{33}}e^{\phi_3 (Y_{12}-Y_{21})}\, .\label{L_SO2diag}
\end{equation}
The resulting scalar potential is rather complicated and will not be needed in the following analysis. Therefore, we refrain from giving it here. 
\\
\indent To preserve some supersymmetry, we perform a topological twist by turning on the following gauge fields
\begin{equation}
A^3=aF'_\kappa(\theta)d\phi+\tilde{a}\tilde{F}'_{\tilde{\kappa}}(\tilde{\theta})d\tilde{\phi}\qquad \textrm{and}\qquad A^6=bF'_\kappa(\theta)d\phi+\tilde{b}\tilde{F}'_{\tilde{\kappa}}(\tilde{\theta})d\tilde{\phi}\label{SO2diag_gauge_field}
\end{equation}
with the condition $g_1A^3=g_2A^6$ or equivalently $g_1a=g_2b$ and $g_1\tilde{a}=g_2\tilde{b}$ implementing the $SO(2)_{\textrm{diag}}$ subgroup. We also note the corresponding field strength tensors
\begin{eqnarray}
& &F^3=-\kappa aF_\kappa(\theta)d\theta\wedge d\phi-\tilde{\kappa}\tilde{a}\tilde{F}_{\tilde{\kappa}}(\tilde{\theta})d\tilde{\theta}\wedge d\tilde{\phi},\nonumber \\
& &F^6=-\kappa bF_\kappa(\theta)d\theta\wedge d\phi-\tilde{\kappa}\tilde{b}\tilde{F}_{\tilde{\kappa}}(\tilde{\theta})d\tilde{\theta}\wedge d\tilde{\phi}\, .
\end{eqnarray}
The composite connection can be straightforwardly computed to be 
\begin{equation}
\Omega_{\mu r0}=0\qquad \textrm{and}\qquad \Omega_{\mu st}=g_1A^3_\mu (\delta_{s2}\delta_{t1}-\delta_{s1}\delta_{t2})
\end{equation}
which lead to
\begin{equation}
Q_{\mu AB}=\frac{i}{2}\sigma_{rAB}
\left[\frac{1}{2}\epsilon^{rst}\Omega_{\mu st}-i\gamma_7
\Omega_{\mu r0}\right]=-\frac{i}{2}g_1A^3_\mu \sigma^3_{AB}\, .\label{SO2diag_composite}
\end{equation}
We also note that only the parts involving gauge fields in the composite connection have been given above. There are additional contributions along the $r$-direction due to non-vanishing scalars $\phi_0$ and $\phi_3$. However, these do not affect the topological twist, so we have omitted them.
\\
\indent We now consider relevant parts of the supersymmetry transformation $\delta \psi_{\hat{\phi}A}$ and $\delta \psi_{\hat{\tilde{\phi}}A}$. For $\delta \psi_{\hat{\phi}A}$, we find
\begin{equation}
0=\frac{1}{2}\frac{F'_\kappa(\theta)}{F_\kappa(\theta)}e^{-h}\gamma_{\hat{\phi}\hat{\theta}}\epsilon_A-\frac{i}{2}g_1a\frac{F'_\kappa(\theta)}{F_\kappa(\theta)}e^{-h}\sigma^3_{AB}\epsilon^B+\ldots
\end{equation}
with $\ldots$ refers to other terms independent of $\theta$ and $\phi$. By imposing a projector and a twist condition
\begin{equation}
\gamma_{\hat{\theta}\hat{\phi}}\epsilon_A=\mp i\sigma^3_{AB}\epsilon^B\qquad \textrm{and}\qquad g_1a=\pm 1\label{SO2diag_twist}
\end{equation}
we can cancel the internal spin connection along $\Sigma$. As a result, $\delta \psi_{\hat{\theta}A}$ and $\delta \psi_{\hat{\phi}A}$ conditions reduce to the same BPS equation for the warp factor $h(r)$ as expected. A similar analysis for $\delta \psi_{\hat{\tilde{\phi}}A}$ gives
\begin{equation}
\gamma_{\hat{\tilde{\theta}}\hat{\tilde{\phi}}}\epsilon_A=\mp i\sigma^3_{AB}\epsilon^B\qquad \textrm{and}\qquad g_1\tilde{a}=\pm 1\, .
\end{equation}
We also note that using the definition of $\gamma_7=i\gamma^{\hat{t}}\gamma^{\hat{r}}\gamma^{\hat{\theta}}\gamma^{\hat{\phi}}
\gamma^{\hat{\tilde{\theta}}}\gamma^{\hat{\tilde{\phi}}}$, the two projectors imply
\begin{equation}
\gamma^{\hat{t}\hat{r}}\epsilon_A=i\gamma_7\epsilon_A\, .
\end{equation}
\indent For the gauge fields given in \eqref{SO2diag_gauge_field}, we cannot consistently set the two-form field to zero. Using the field equation \eqref{B_eq} given in the appendix, we find that $H_{\mu\nu\rho}=0$ and only $B_{\hat{t}\hat{r}}$ component is non-vanishing and given by
\begin{equation}
m^2e^{-2\sigma}\mc{N}_{00}B^{\hat{t}\hat{r}}=-\frac{1}{16}\epsilon^{\hat{t}\hat{r}\hat{\rho}\hat{\sigma}\hat{\lambda}\hat{\tau}}\eta_{\Lambda\Sigma}F^\Lambda_{\hat{\rho}\hat{\sigma}}
F^\Sigma_{\hat{\lambda}\hat{\tau}}\, .
\end{equation}  
With $\epsilon^{\hat{t}\hat{r}\hat{\theta}\hat{\phi}\hat{\tilde{\theta}}\hat{\tilde{\phi}}}=1$, we find
\begin{equation}
B_{\hat{t}\hat{r}}=\frac{1}{8}\frac{\kappa\tilde{\kappa}(a\tilde{a}-b\tilde{b})}{m^2\mc{N}_{00}}e^{2\sigma-2h-2\tilde{h}}\, .
\end{equation}
\indent To obtain the BPS equations, we also need to impose a projector involving $\gamma^{\hat{r}}$ 
\begin{equation}
\gamma^{\hat{r}}\epsilon_A=-\epsilon_A\, .\label{gamma_r_proj}
\end{equation}
The sign choice is chosen such that an $AdS_6$ vacuum appears at $r\rightarrow \infty$ in the solutions. It turns out that consistency of the BPS equations requires $\phi_0=0$ as in the solutions studied in \cite{AdS6_BH_Zaffaroni} and \cite{AdS6_BH_Minwoo}. With all these, we find the following BPS equations
\begin{eqnarray}
\phi'_1&=&-e^{\sigma}\textrm{sech}2\phi_3\sinh2\phi_1(g_1\cosh\phi_2-g_2\sinh\phi_2),\\
\phi'_2&=&-e^\sigma\left[g_2\cosh\phi_2+g_1\sinh\phi_2+\cosh2\phi_1\cosh2\phi_3(g_1\sinh\phi_2-g_2\cosh\phi_2)\right]\nonumber \\
& &+\frac{1}{2}e^{-\sigma-2h-2\tilde{h}}\left[\kappa e^{2\tilde{h}}(b\cosh\phi_2-a\sinh\phi_2)-\tilde{\kappa}e^{2h}(\tilde{b}\cosh\phi_2-\tilde{a}\sinh\phi_2)\right],
\nonumber \\
\\
\phi'_3&=&-e^\sigma\cosh2\phi_1\sinh2\phi_3(g_1\cosh\phi_2-g_2\sinh\phi_2),
\end{eqnarray}
\begin{eqnarray}
\sigma'&=&-\frac{1}{4}e^\sigma\left[g_1\cosh\phi_2+g_2\sinh\phi_2+\cosh2\phi_1\cosh2\phi_3(g_1\cosh\phi_2-g_2\sinh\phi_2)\right]\nonumber\\
& &+\frac{1}{8}e^{-\sigma-2h-2\tilde{h}}\left[\kappa e^{2\tilde{h}}(a\cosh\phi_2-b\sinh\phi_2)+\tilde{\kappa}e^{2h}(\tilde{a}\cosh\phi_2-\tilde{b}\sinh\phi_2)\right]\nonumber \\ 
& &+\frac{3}{2}me^{-3\sigma}-\frac{1}{32m}e^{\sigma-2h-2\tilde{h}}\kappa\tilde{\kappa}(b\tilde{b}-a\tilde{a})\\
h'&=&\frac{1}{4}e^{\sigma}\left[g_1\cosh\phi_2+g_2\sinh\phi_2+\cosh2\phi_1\cosh2\phi_3(g_1\cosh\phi_2-g_2\sinh\phi_2)\right]\nonumber \\
& &+\frac{1}{8}e^{-\sigma-2h-2\tilde{h}}\left[3\kappa e^{2\tilde{h}}(a\cosh\phi_2-b\sinh\phi_2)-\tilde{\kappa}e^{2h}(\tilde{a}\cosh\phi_2-\tilde{b}\sinh\phi_2)\right]\nonumber \\
& &+\frac{1}{2}me^{-3\sigma}+\frac{1}{32m}e^{\sigma-2h-2\tilde{h}}\kappa\tilde{\kappa}(b\tilde{b}-a\tilde{a})\\
\tilde{h}'&=&\frac{1}{4}e^{\sigma}\left[g_1\cosh\phi_2+g_2\sinh\phi_2+\cosh2\phi_1\cosh2\phi_3(g_1\cosh\phi_2-g_2\sinh\phi_2)\right]\nonumber \\
& &+\frac{1}{8}e^{-\sigma-2h-2\tilde{h}}\left[3\tilde{\kappa}e^{2h}(\tilde{a}\cosh\phi_2-\tilde{b}\sinh\phi_2)-\kappa e^{2\tilde{h}}(a\cosh\phi_2-b\sinh\phi_2)\right]\nonumber \\
& &+\frac{1}{2}me^{-3\sigma}+\frac{1}{32m}e^{\sigma-2h-2\tilde{h}}\kappa\tilde{\kappa}(b\tilde{b}-a\tilde{a})\\
f'&=&\frac{1}{4}e^{\sigma}\left[g_1\cosh\phi_2+g_2\sinh\phi_2+\cosh2\phi_1\cosh2\phi_3(g_1\cosh\phi_2-g_2\sinh\phi_2)\right]\nonumber \\
& &-\frac{1}{8}e^{-\sigma-2h-2\tilde{h}}\left[\kappa e^{2\tilde{h}}(a\cosh\phi_2-b\sinh\phi_2)+\tilde{\kappa}e^{2h}(\tilde{a}\cosh\phi_2-\tilde{b}\sinh\phi_2)\right]\nonumber \\
& &+\frac{1}{2}me^{-3\sigma}-\frac{3}{32m}e^{\sigma-2h-2\tilde{h}}\kappa\tilde{\kappa}(b\tilde{b}-a\tilde{a})\, .
\end{eqnarray}
In deriving these equations, we have chosen the upper sign choice for the conditions given in \eqref{SO2diag_twist} for definiteness. It can also be verified that these equations are compatible with the second-order field equations. For large $r$ with $f\sim h\sim \tilde{h}\sim r$, these equations admit $AdS_6$ vacua given in \eqref{SO4_AdS6} and \eqref{SO3_AdS6} as asymptotic solutions.  
\\
\indent We note that for $\phi_1=\phi_3=0$ and without the $SO(2)_{\textrm{diag}}$ condition among the gauge fields $A^3$ and $A^6$, we recover the BPS equations given in \cite{AdS6_BH_Zaffaroni} and \cite{AdS6_BH_Minwoo} for $SO(2)\times SO(2)$ symmetric black holes up to some notational and conventional differences. However, in the present case, the magnetic charges $b$ and $\tilde{b}$ are not independent but related to $a$ and $\tilde{a}$ via the relations $g_2b=g_1a$ and $g_2\tilde{b}=g_1\tilde{a}$. The solutions to these equations preserve $\frac{1}{8}$ supersymmetry or $2$ supercharges due to three independent projectors involving $\gamma_{\hat{\theta}\hat{\phi}}$, $\gamma_{\hat{\tilde{\theta}}\hat{\tilde{\phi}}}$, and $\gamma^{\hat{r}}$. 
\\
\indent At the horizon given by an $AdS_2\times \Sigma\times \widetilde{\Sigma}$ critical point, we have the conditions
\begin{equation}
\sigma'=h'=\tilde{h}'=\phi_i'=0,\qquad i=1,2,3,\qquad\textrm{and}\qquad  f'=\frac{1}{\ell} 
\end{equation}
with an $AdS_2$ radius $\ell$. The constant scalars imply that the $\gamma^{\hat{r}}$ projector is not needed in the BPS equations, and the $AdS_2\times \Sigma\times \widetilde{\Sigma}$ solutions are $\frac{1}{4}$-BPS preserving four supercharges.

\subsection{$AdS_2\times \Sigma\times \widetilde{\Sigma}$ vacua}
We now look for possible $AdS_2\times \Sigma\times \widetilde{\Sigma}$ vacua from the above BPS equations. The first solution is given by
\begin{itemize}
\item $AdS_2$ critical point I:
\begin{eqnarray}
& &\phi_1=\phi_3=0,\nonumber\\
& & h=\frac{1}{2}\ln\left[-\frac{e^{2\sigma}\kappa(a\cosh\phi_2-b\sinh\phi_2)}{4m}\right],\nonumber \\
& & \tilde{h}=\frac{1}{2}\ln\left[-\frac{e^{2\sigma}\tilde{\kappa}(\tilde{a}\cosh\phi_2-\tilde{b}\sinh\phi_2)}{4m}\right],\nonumber \\
& &\sigma=\frac{1}{4}\ln\left[\frac{m[3(b\tilde{b}-a\tilde{a})-(a\tilde{a}+b\tilde{b})\cosh2\phi_2+(a\tilde{b}+b\tilde{a})\sinh2\phi_2]}{2g_1\cosh^3\phi_2(a-b\tanh\phi_2)(\tilde{b}\tanh\phi_2-\tilde{a})}\right],\nonumber\\
& &\phi_2=\frac{1}{2}\ln\left[\frac{2\Phi^{\frac{2}{3}}+(3g_1+g_2)\Phi^{\frac{1}{3}}+2g_1(3g_1+2g_2)}{(g_2-g_1)\Phi^{\frac{1}{3}}}\right],\nonumber \\
& &\frac{1}{\ell}=\left[\frac{8mg_1^3\cosh\phi_2[3(b\tilde{b}-a\tilde{a})-(a\tilde{a}+b\tilde{b})\cosh2\phi_2+(a\tilde{b}+b\tilde{a})\sinh2\phi_2]}{(a-b\tanh\phi_2)(\tilde{b}\tanh\phi_2-\tilde{a})}\right]^{\frac{1}{4}}\nonumber \\
\label{AdS2_Sigma_Sigma1}
\end{eqnarray}
with 
\begin{equation}
\Phi=5g_1^3+5g_1^2g_2+g_1g_2^2+g_1(g_1+g_2)\sqrt{g_2^2-2g_1^2}\, .
\end{equation}
\end{itemize}
To simplify some expressions, we have used the relations $g_1a=g_2b$ and $g_1\tilde{a}=g_2\tilde{b}$ which lead to a less symmetric appearance of $(a,\tilde{a})$ and $(b,\tilde{b})$. A straightforward analysis shows that only $\kappa=\tilde{\kappa}=-1$ leads to valid $AdS_2\times \Sigma\times \widetilde{\Sigma}$ solutions. 
\\
\indent There is also another $AdS_2\times \Sigma\times \widetilde{\Sigma}$ solution of the form
\begin{itemize}
\item $AdS_2$ critical point II:
\begin{eqnarray}
& &\phi_2=\frac{1}{2}\ln\left[\frac{g_2+g_1}{g_2-g_1}\right],\nonumber\\
& & h=\frac{1}{2}\ln\left[-\frac{e^{2\sigma}\kappa(ag_2-g_1b)}{4m\sqrt{g^2_2-g_1^2}}\right],\quad \tilde{h}=\frac{1}{2}\ln\left[-\frac{e^{2\sigma}\tilde{\kappa}(\tilde{a}g_2-\tilde{b}g_1)}{4m\sqrt{g^2_2-g_1^2}}\right],\nonumber \\
& &\sigma=\frac{1}{4}\ln\left[\frac{m[a\tilde{a}(2g_2^2-g_1^2)+b\tilde{b}(g_1^2-g_2^2)-g_1g_2(a\tilde{b}+\tilde{a}b)]\sqrt{g_2^2-g_1^2}}{g_1g_2(ag_2-bg_1)(\tilde{a}g_2-\tilde{b}g_1)}\right],\nonumber\\
& &\phi_1=\frac{1}{2}\ln\left[\frac{2(g_1^2+g_2^2)+\sqrt{2}\sqrt{g_1^4+6g_1^2g_2^2+g_2^4-(g_2^2-g_1^2)\cosh4\phi_3}}{2(g_2^2-g_1^2)\cosh2\phi_3}\right],
\nonumber \\
& &\frac{1}{\ell}=\left[\frac{16mg_1^3g_2^3[a\tilde{a}(2g_2^2-g_1^2)+b\tilde{b}(g_2^2-2g_1^2)-g_1g_2(a\tilde{b}+\tilde{a}b)]}{(g_2^2-g_1^2)^{\frac{3}{2}}(ag_2-bg_1)(\tilde{a}g_2-\tilde{b}g_1)}\right]^{\frac{1}{4}}\, \, \,
 \label{AdS2_Sigma_Sigma2}
\end{eqnarray}
with $\phi_3$ being a constant. We also note that apart from $\phi_1$, all other functions including the $AdS_2$ radius do not depend on the value of $\phi_3$ at the critical point. For a particular value of $\phi_3=0$, we find
\begin{equation}
\phi_1=\frac{1}{2}\ln\left[\frac{g_2+g_1}{g_2-g_1}\right].
\end{equation}
\end{itemize}
As in the case of critical point I, it turns out that only $\kappa=\tilde{\kappa}=-1$ leads to viable solutions. Therefore, there are only black hole solutions with $AdS_2\times H^2\times H^2$ near horizon geometries.

\subsection{Numerical black hole solutions}
\begin{figure}
         \centering
               \begin{subfigure}[b]{0.45\textwidth}
                 \includegraphics[width=\textwidth]{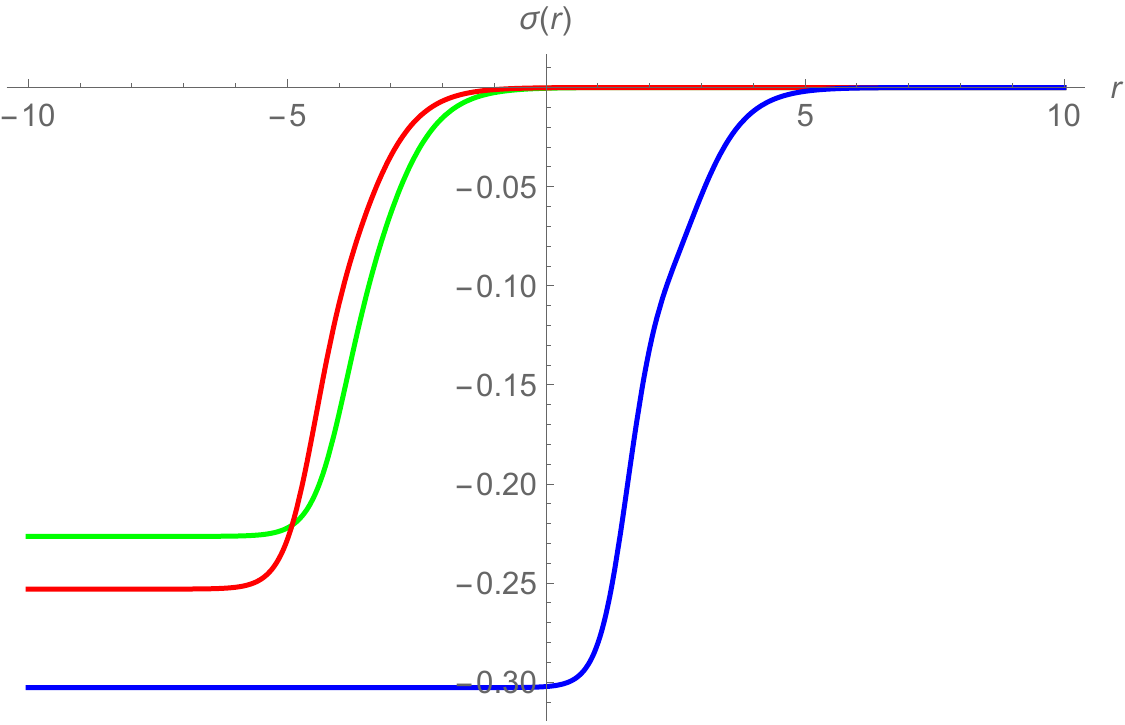}
                 \caption{Solutions for $\sigma(r)$}
         \end{subfigure}
         \begin{subfigure}[b]{0.45\textwidth}
                 \includegraphics[width=\textwidth]{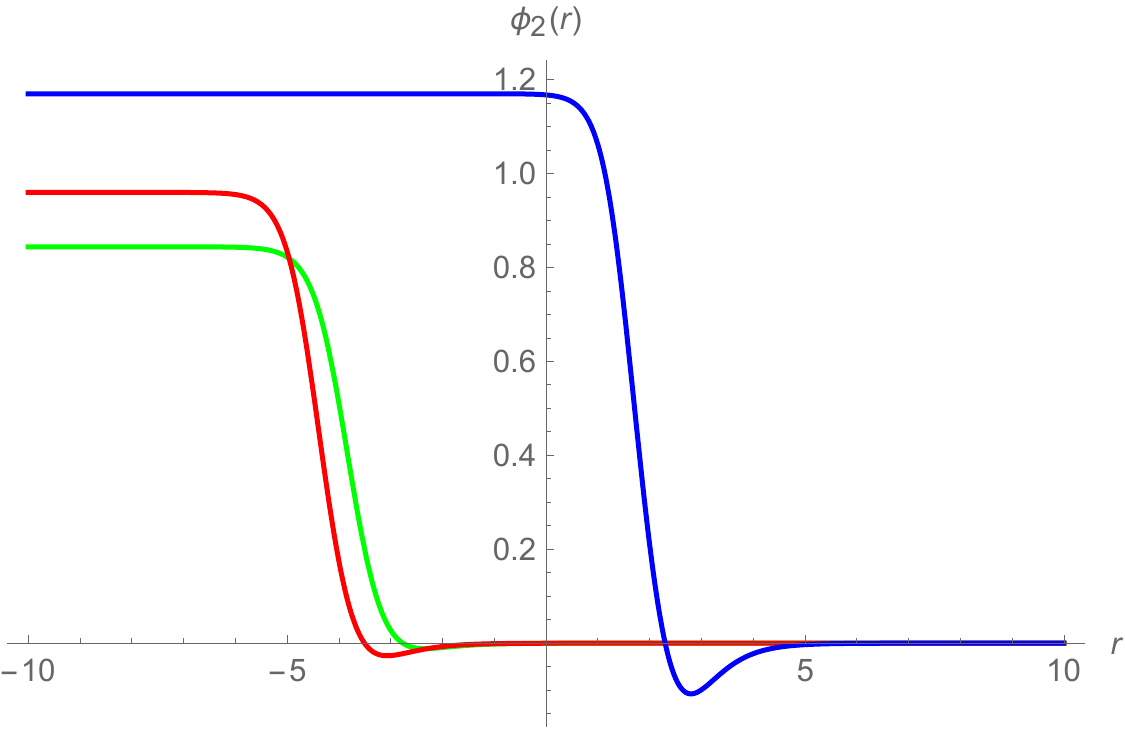}
                 \caption{Solutions for $\phi_2(r)$}
         \end{subfigure}\\
          \begin{subfigure}[b]{0.45\textwidth}
                 \includegraphics[width=\textwidth]{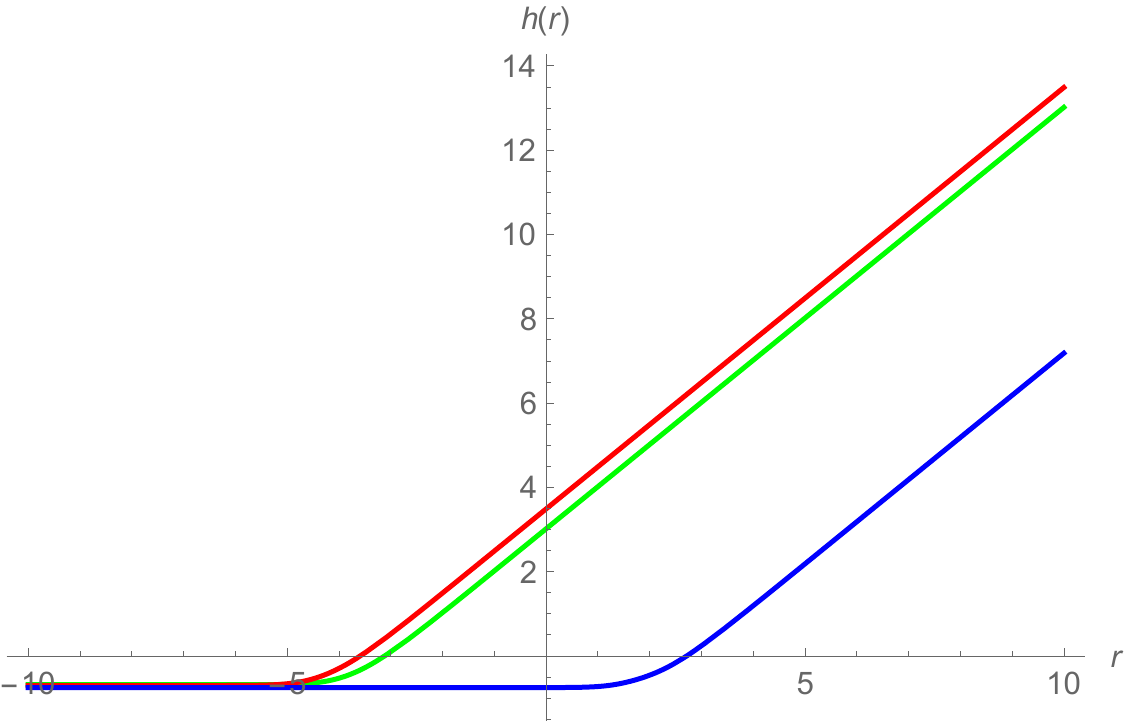}
                 \caption{Solutions for $h(r)$}
         \end{subfigure}
          \begin{subfigure}[b]{0.45\textwidth}
                 \includegraphics[width=\textwidth]{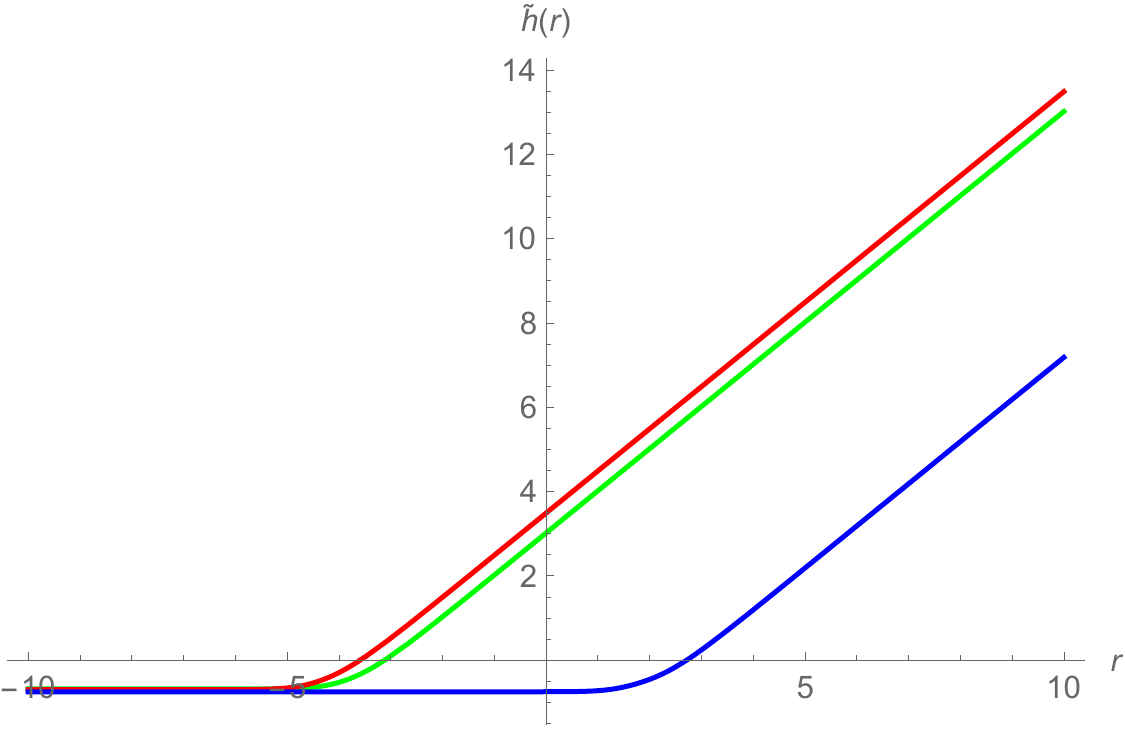}
                 \caption{Solutions for $\tilde{h}(r)$}
         \end{subfigure}\\
          \begin{subfigure}[b]{0.45\textwidth}
                 \includegraphics[width=\textwidth]{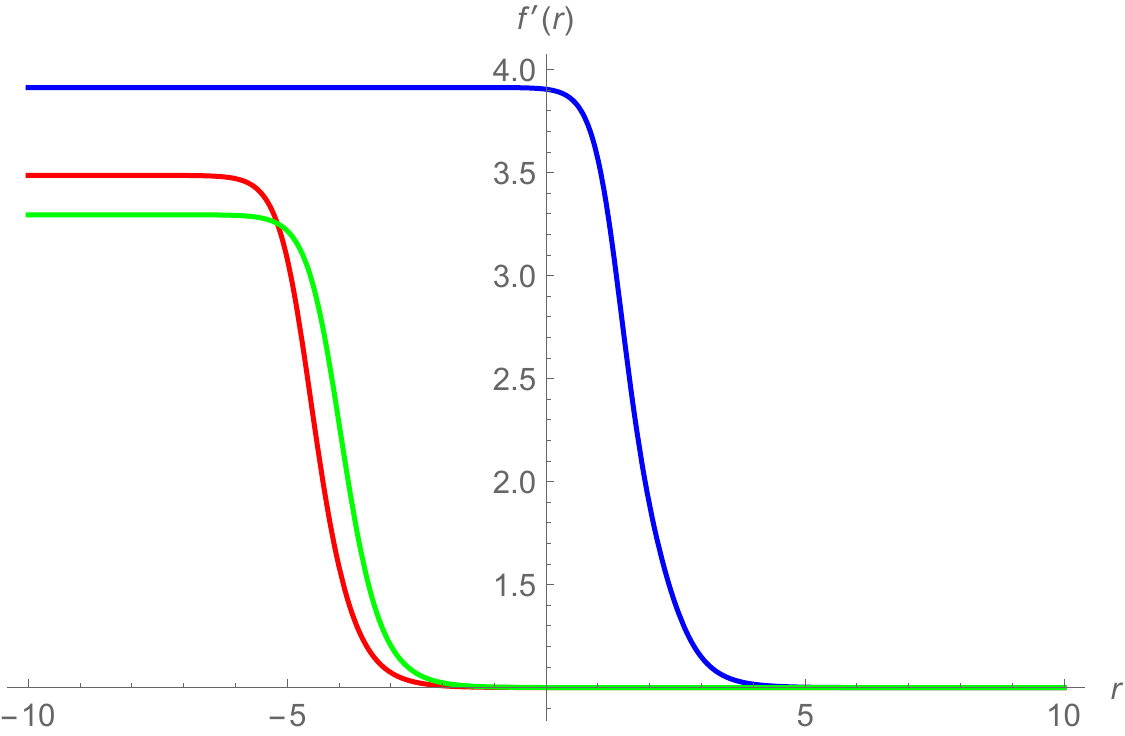}
                 \caption{Solutions for $f'(r)$}
         \end{subfigure}
\caption{Supersymmetric $AdS_6$ black holes interpolating between $AdS_6$ vacuum $i$ and the near horizon geometry $AdS_2\times H^2\times H^2$ (critical point I) for $g_2=4$ (blue), $g_2=6$ (red), $g_2=8$ (green).}\label{fig1}
 \end{figure}

\begin{figure}
         \centering
               \begin{subfigure}[b]{0.45\textwidth}
                 \includegraphics[width=\textwidth]{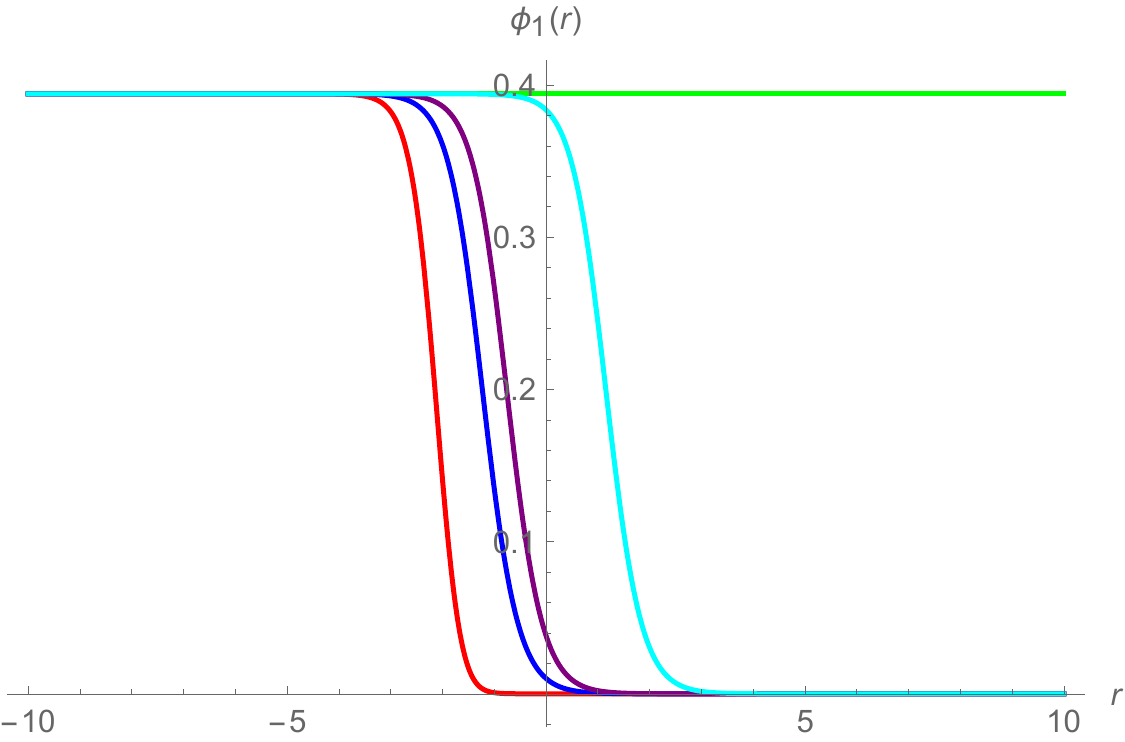}
                 \caption{Solutions for $\phi_1(r)$}
         \end{subfigure}
         \begin{subfigure}[b]{0.45\textwidth}
                 \includegraphics[width=\textwidth]{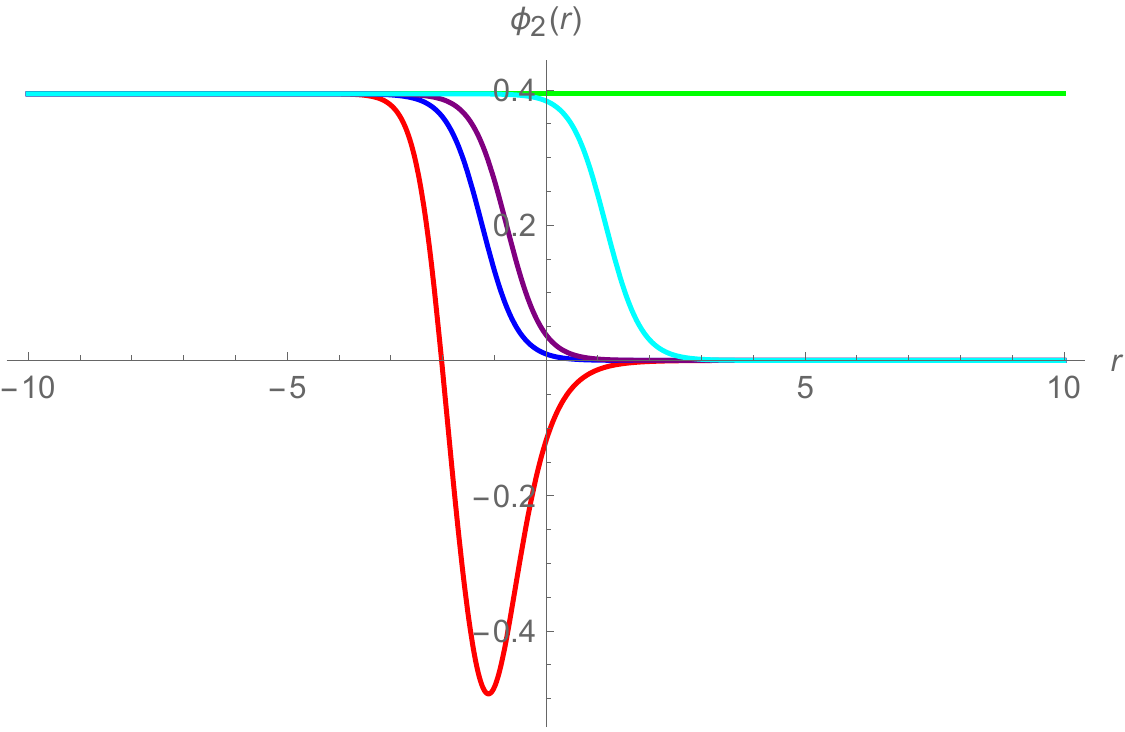}
                 \caption{Solutions for $\phi_2(r)$}
         \end{subfigure}\\
          \begin{subfigure}[b]{0.45\textwidth}
                 \includegraphics[width=\textwidth]{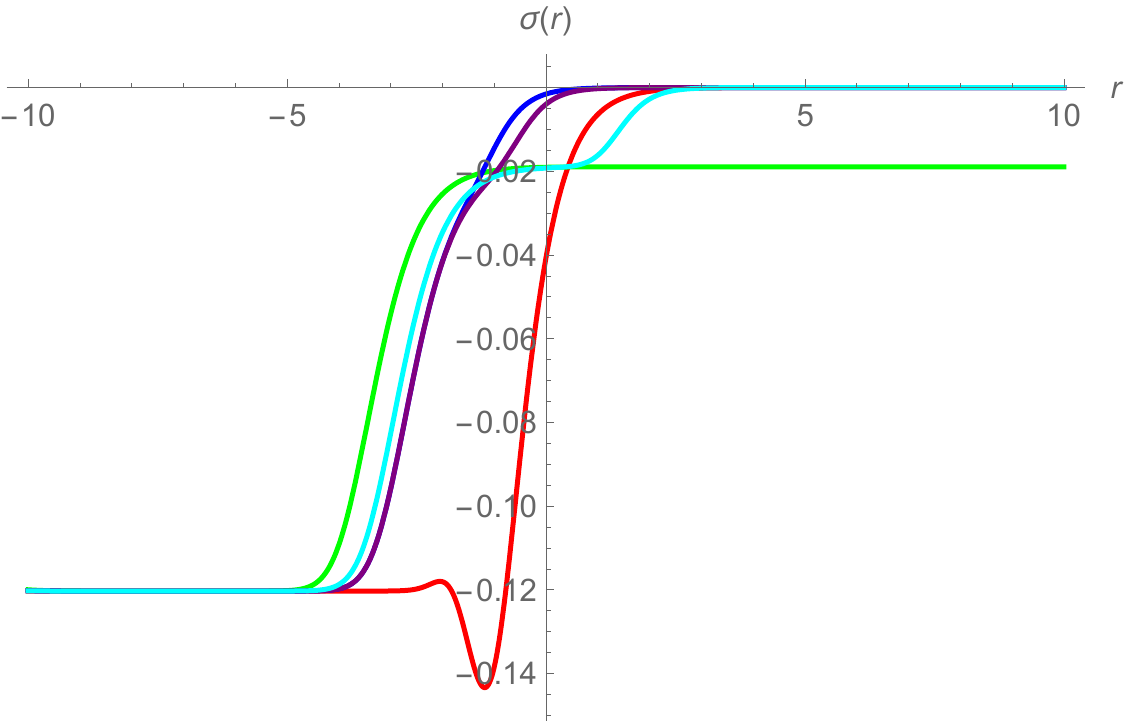}
                 \caption{Solutions for $\sigma(r)$}
         \end{subfigure}
          \begin{subfigure}[b]{0.45\textwidth}
                 \includegraphics[width=\textwidth]{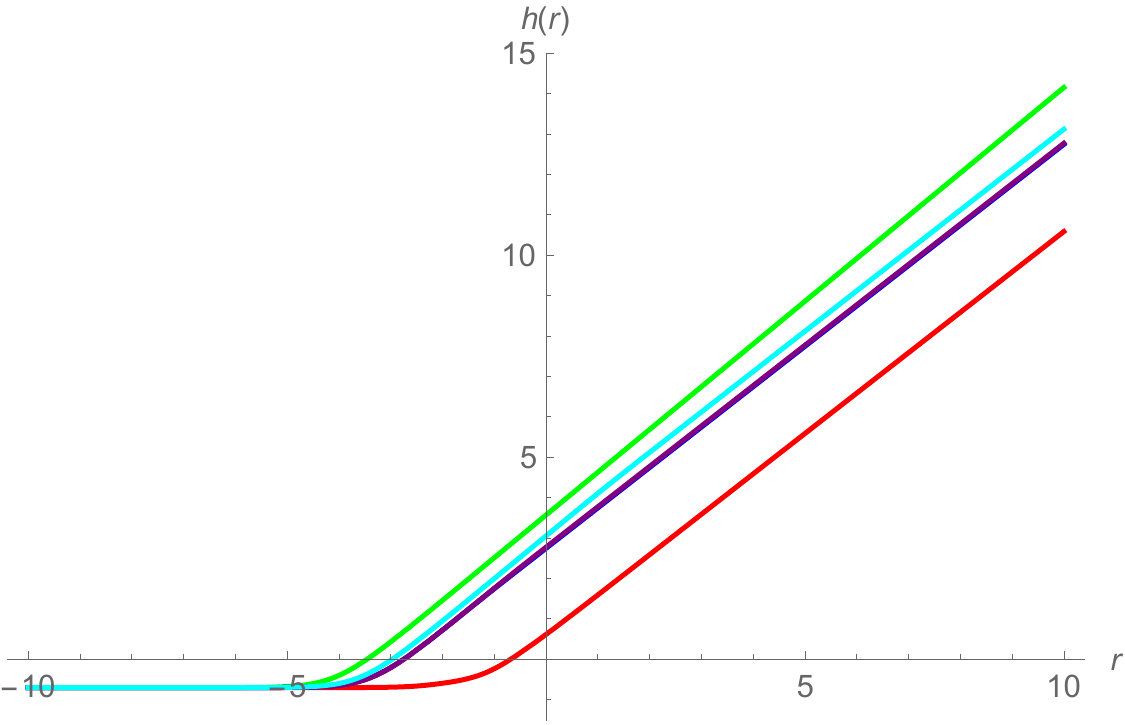}
                 \caption{Solutions for $h(r)$}
         \end{subfigure}\\
                   \begin{subfigure}[b]{0.45\textwidth}
                 \includegraphics[width=\textwidth]{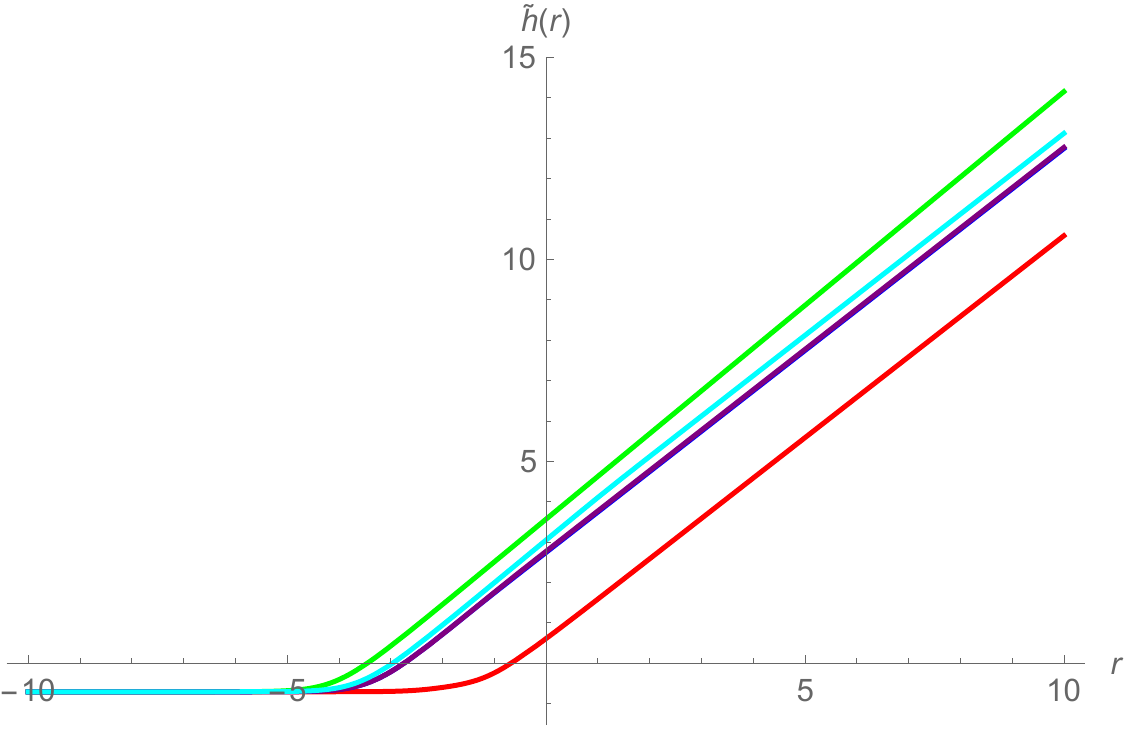}
                 \caption{Solutions for $\tilde{h}(r)$}
           \end{subfigure}      
          \begin{subfigure}[b]{0.45\textwidth}
                 \includegraphics[width=\textwidth]{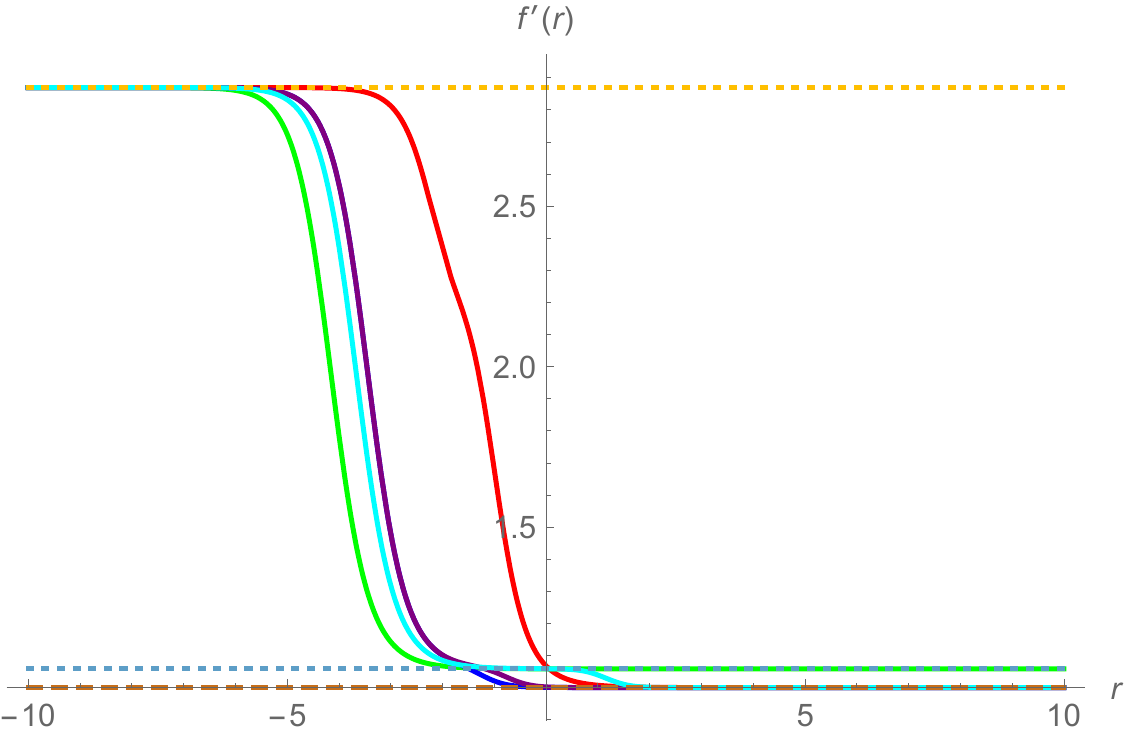}
                 \caption{Solutions for $f'(r)$}
         \end{subfigure}
\caption{Supersymmetric $AdS_6$ black holes interpolating between $AdS_6$ vacua $i$ and $ii$ and the near horizon geometry $AdS_2\times H^2\times H^2$ (critical point II) for $g_2$=4. The red and green curves represent respectively solutions that flows directly from $AdS_6$ critical points $i$ and $ii$ to $AdS_2\times H^2\times H^2$ vacuum II. Solutions interpolating between $AdS_6$ critical point $i$ and $AdS_2\times H^2\times H^2$ vacuum II that flow very close to $AdS_6$ vacuum $ii$ are given by blue, purple, and cyan curves. Note also that in subfigures (d), (e), and (f), solutions represented by blue and purple lines are very close to each other rendering the blue line invisible.}\label{fig2}
\end{figure}

The BPS equations are too complicated to be solved analytically, so we will numerically find the black hole solutions interpolating between asymptotically locally $AdS_6$ vacua and near horizon $AdS_2\times H^2\times H^2$ geometries. According to the AdS/CFT correspondence, the solutions also describe RG flows across dimensions from five-dimensional SCFTs in the UV to superconformal quantum mechanics in the IR. 
\\
\indent We begin with solutions flowing to $AdS_2$ critical point I. We will choose the values of various parameters as follows
\begin{equation}
g_1=3m\qquad \textrm{and}\qquad m=\frac{1}{2}\, .
\end{equation}
With these values, the supersymmetric $AdS_6$ critical point with $SO(3)\times SO(3)$ symmetry given in \eqref{SO4_AdS6} has unit radius. We will also set $\phi_1=\phi_3=0$ along the entire solutions. Examples of solutions with different values of $g_2$ are shown in figure \ref{fig1}. We can also compute the corresponding black hole entropy by using the relation
\begin{equation}
S_{\textrm{BH}}=\frac{1}{4G_{\textrm{N}}}e^{2h+2\tilde{h}}\textrm{vol}(\Sigma)\textrm{vol}(\widetilde{\Sigma})
\end{equation}
with the volume of a genus-$\mathfrak{g}$ Riemann surface given by
\begin{equation}
\textrm{vol}(\Sigma_{\mathfrak{g}})=2\pi \eta_{\mathfrak{g}},\qquad \eta_{\mathfrak{g}}=\begin{cases}
  2|\mathfrak{g}-1|,  &\quad  \textrm{for}\quad \mathfrak{g}\neq 1 \\
  1,  & \quad \textrm{for}\quad \mathfrak{g}=1
\end{cases}\, .
\end{equation}
For the present case of black hole with $H^2\times H^2$ horizon, we find 
\begin{equation}
S_{\textrm{BH}}=\frac{\pi^2\eta_{\mathfrak{g}}\eta_{\tilde{\mathfrak{g}}}\left[(a\tilde{a}+b\tilde{b})\cosh\phi_2+(a\tilde{a}-2b\tilde{b})\textrm{sech}\phi_2-(\tilde{a}b
+a\tilde{b})\sinh\phi_2\right]}{16g_1mG_{\textrm{N}}}
\end{equation}    
with the value of $\phi_2$ at the horizon given in \eqref{AdS2_Sigma_Sigma1}.
 \\
 \indent We now move to black hole solutions with the near horizon geometry given by $AdS_2\times H^2\times H^2$ critical point II. By setting $\phi_3=0$ and $g_2=4$ with all other parameters take the same values as in the previous case, we give examples of black hole solutions in figure \ref{fig2}. In this case, there is a solution that flows directly from $AdS_6$ critical point $i$ to $AdS_2\times H^2\times H^2$ vacuum II (red curve) as well as a solution interpolating between $AdS_6$ critical point $ii$ and $AdS_2\times H^2\times H^2$ vacuum II (green curve). There are also solutions interpolating between $AdS_6$ critical point $i$ and $AdS_2\times H^2\times H^2$ vacuum II that flow very close to $AdS_6$ vacuum $ii$ (blue and purple curves). By fine tuning the boundary conditions, we can find a solution interpolating between $AdS_6$ critical points $i$ and $ii$ and $AdS_2\times H^2\times H^2$ vacuum II (cyan curve). In the solutions for $f'(r)$, we have also included the values of $f'(r)$ at various critical points (dashed lines), related to $AdS_6$ and $AdS_2$ radii, to clearly illustrate the interpolations among different vacua. The corresponding black hole entropy is given by
 \begin{equation}
S_{\textrm{BH}}=\frac{\pi^2\eta_{\mathfrak{g}}\eta_{\tilde{\mathfrak{g}}}\left[a\tilde{a}(2g_2^2-g_1^2)+b\tilde{b}(2g_1^2-g_2^2)-(a\tilde{b}+b\tilde{a})g_1g_2\right]}{16g_1g_2m\sqrt{g_2^2-g_1^2}G_{\textrm{N}}}\, .
\end{equation}    

\subsection{Solutions with $SO(2)_R$ twist} 
Another possibility of performing the topological twist is to turn on only $SO(2)_R\subset SO(3)_R$ gauge field. In this case, there are six singlet scalars parametrized by the coset representative
\begin{equation}
L=e^{\varphi_1Y_{01}}e^{\varphi_2Y_{02}}e^{\varphi_3Y_{03}}e^{\phi_1Y_{31}}e^{\phi_2Y_{32}}e^{\phi_3Y_{33}}\, .
\end{equation} 
Similar to the case of $SO(2)_{\textrm{diag}}$ twist, consistency of the BPS equations requires $\varphi_1=\varphi_2=\varphi_3=0$. The analysis of the BPS equations shows that $AdS_2$ critical points exist only for $\phi_1=\phi_2=\phi_3=0$. This gives the same critical point studied in \cite{AdS6_BH_Minwoo1} and \cite{AdS6_BH_Minwoo}. Accordingly, we will not give further detail on this result here to avoid repetition.

\section{Supersymmetric $AdS_6$ black holes with $\mc{M}_4$ horizons}\label{AdS2_M4}
In this section, we consider $AdS_6$ black holes with a near horizon geometry of the form $AdS_2\times \mc{M}_4$ for $\mc{M}_4$ being an Einstein four-manifold. We will consider two types of $\mc{M}_4$ namely a Kahler four-cycle and a Cayley four-cycle. 

\subsection{Black holes with Kahler four-cycle horizon}
\begin{figure}
         \centering
               \begin{subfigure}[b]{0.45\textwidth}
                 \includegraphics[width=\textwidth]{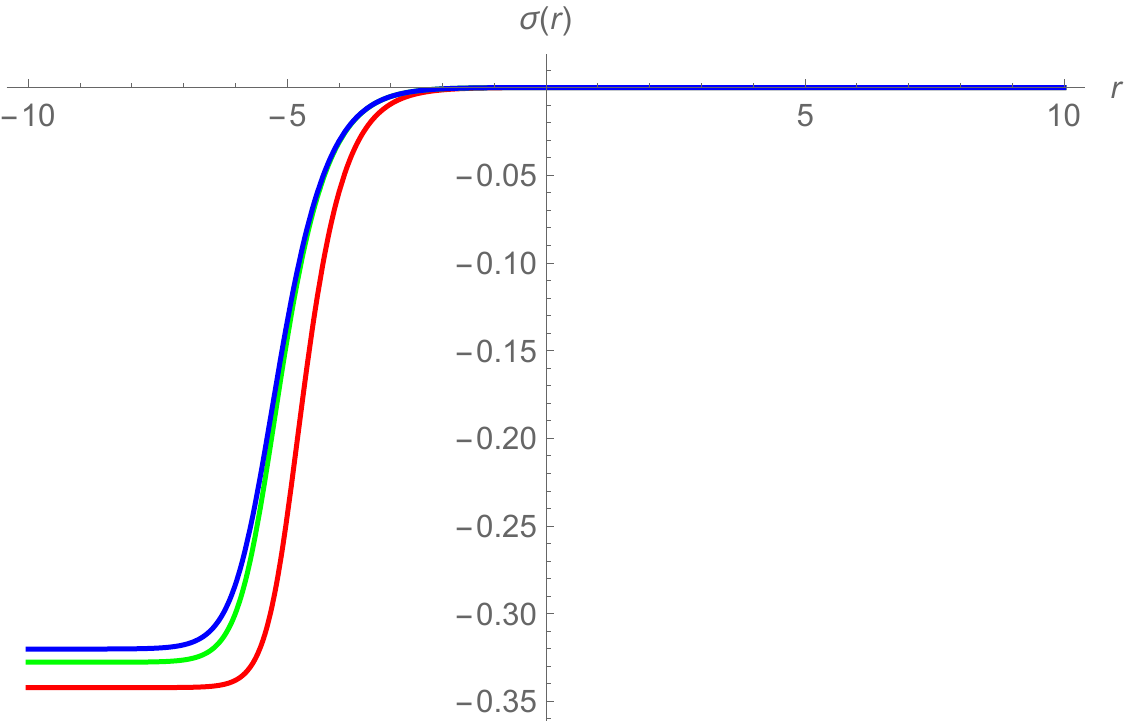}
                 \caption{Solutions for $\sigma(r)$}
         \end{subfigure}
         \begin{subfigure}[b]{0.45\textwidth}
                 \includegraphics[width=\textwidth]{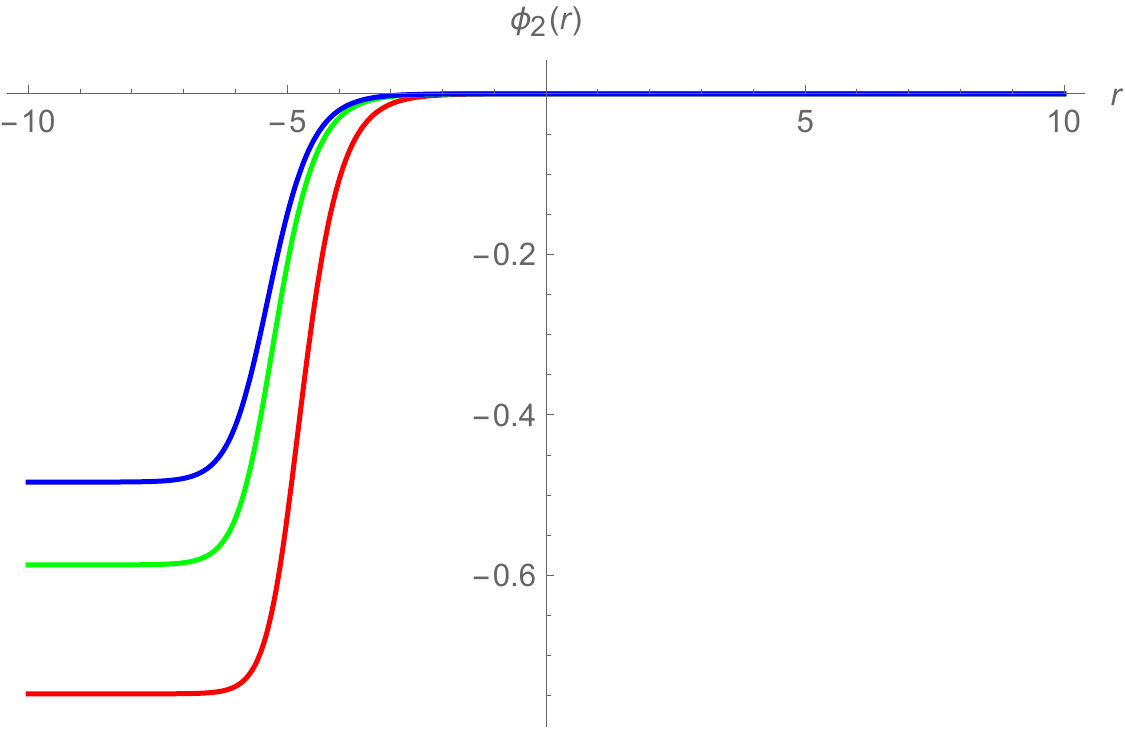}
                 \caption{Solutions for $\phi_2(r)$}
         \end{subfigure}\\
          \begin{subfigure}[b]{0.45\textwidth}
                 \includegraphics[width=\textwidth]{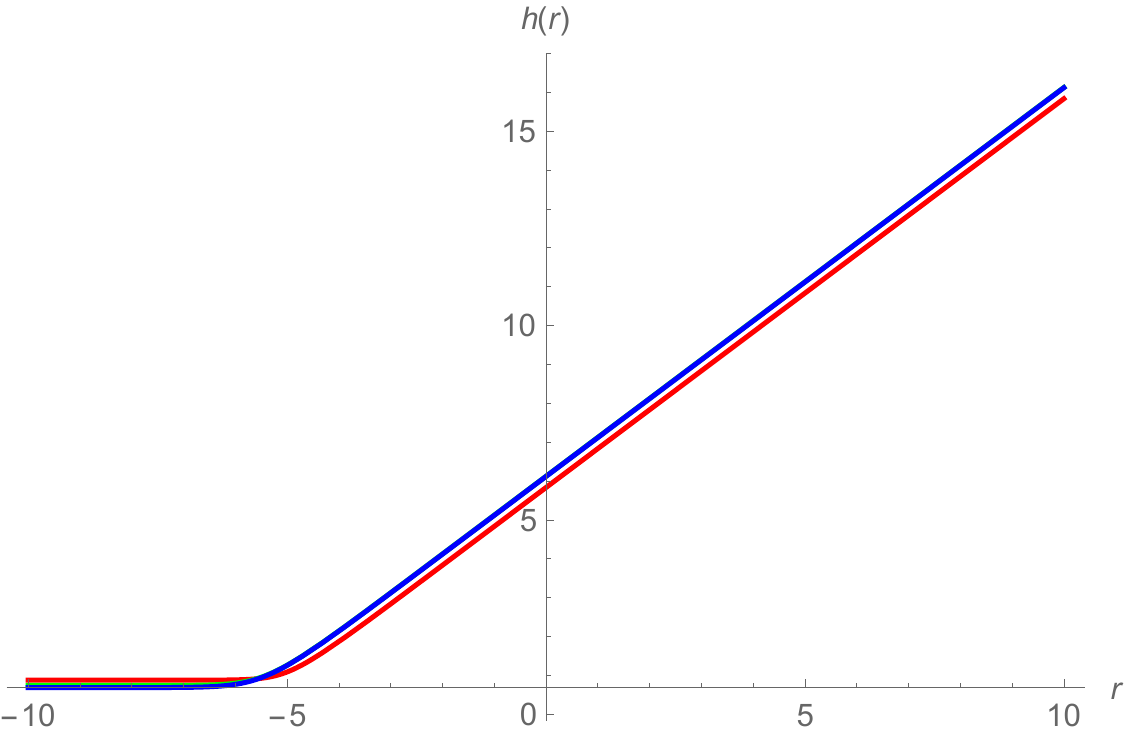}
                 \caption{Solutions for $h(r)$}
         \end{subfigure}
          \begin{subfigure}[b]{0.45\textwidth}
                 \includegraphics[width=\textwidth]{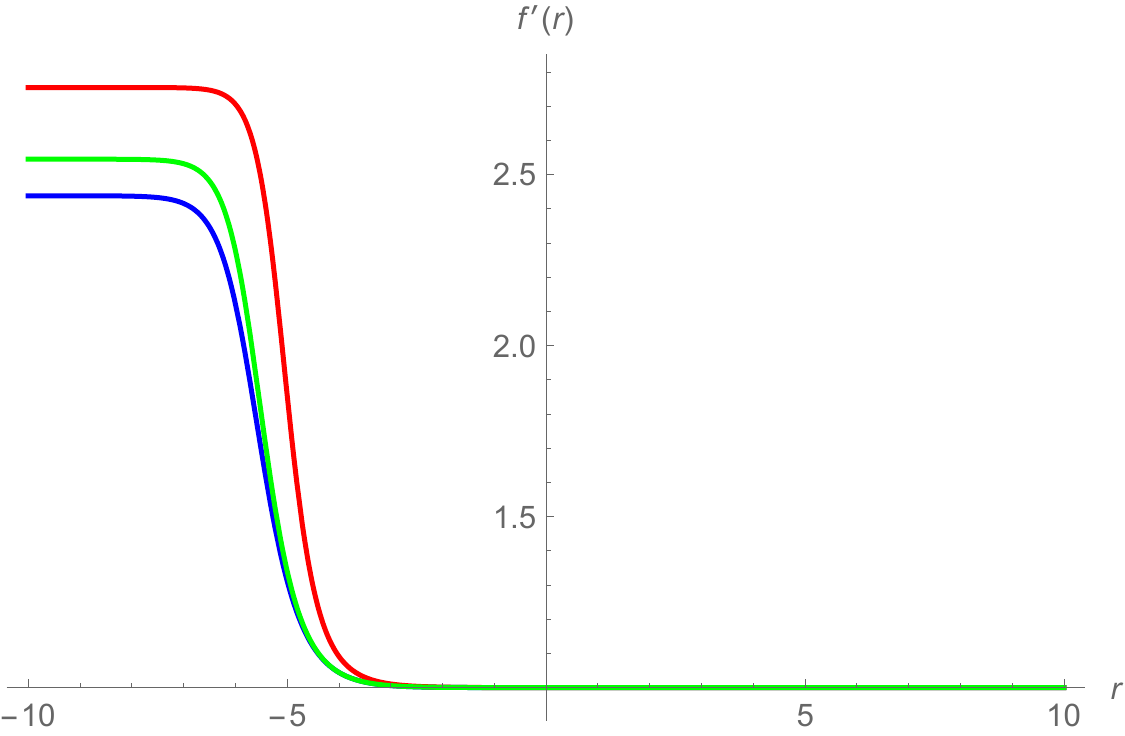}
                 \caption{Solutions for $f'(r)$}
         \end{subfigure}
\caption{Supersymmetric $AdS_6$ black holes interpolating between $AdS_6$ vacuum $i$ and the near horizon geometry $AdS_2\times \mc{M}_4^-$ (critical point III) for $g_2=2$ (red), $g_2=4$ (green), $g_2=6$ (blue).}\label{fig3}
 \end{figure}

\begin{figure}
         \centering
               \begin{subfigure}[b]{0.45\textwidth}
                 \includegraphics[width=\textwidth]{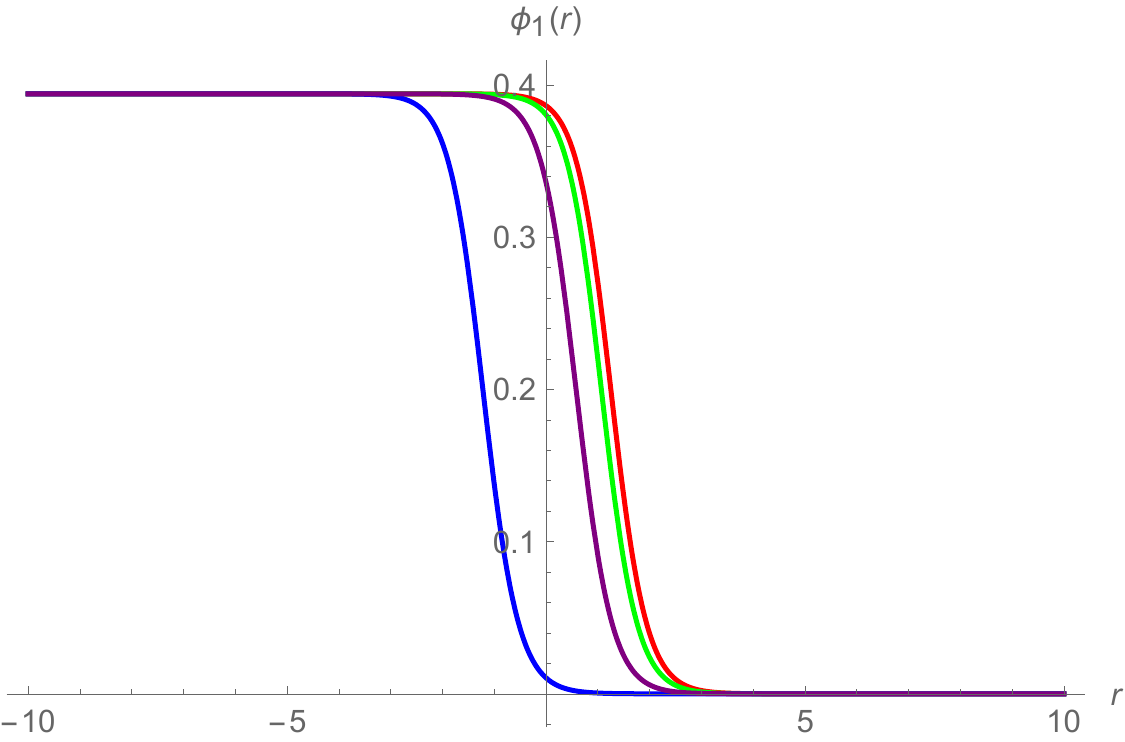}
                 \caption{Solutions for $\phi_1(r)$}
         \end{subfigure}
         \begin{subfigure}[b]{0.45\textwidth}
                 \includegraphics[width=\textwidth]{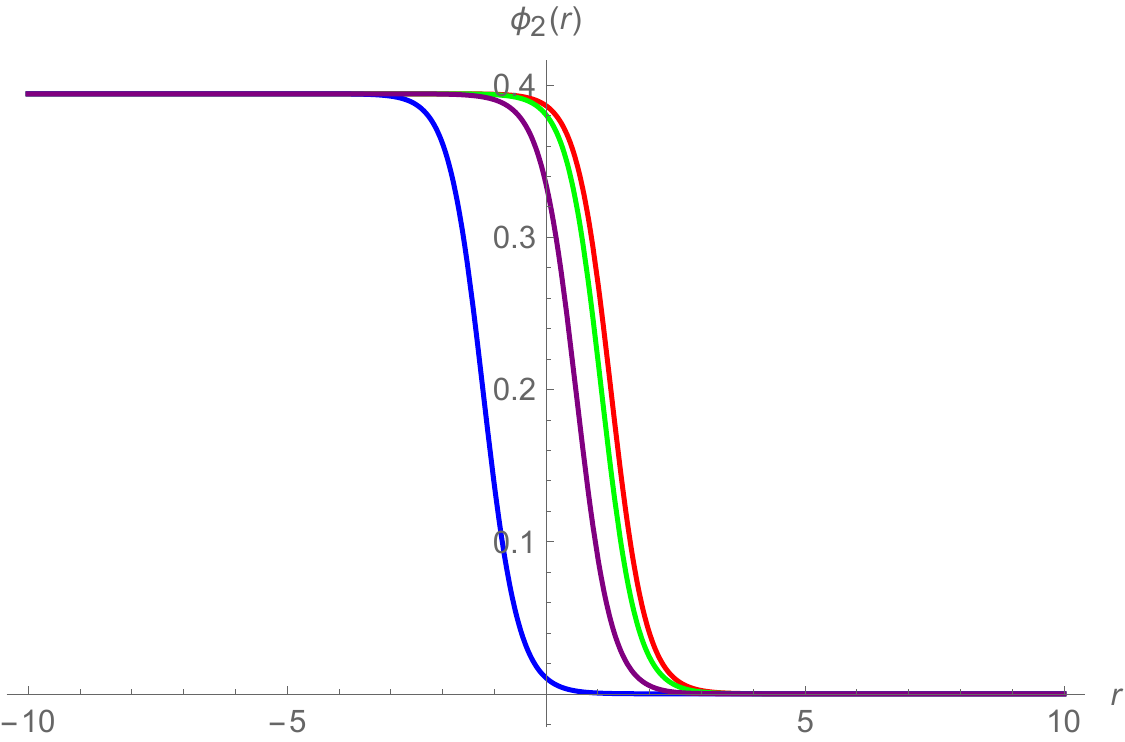}
                 \caption{Solutions for $\phi_2(r)$}
         \end{subfigure}\\
          \begin{subfigure}[b]{0.45\textwidth}
                 \includegraphics[width=\textwidth]{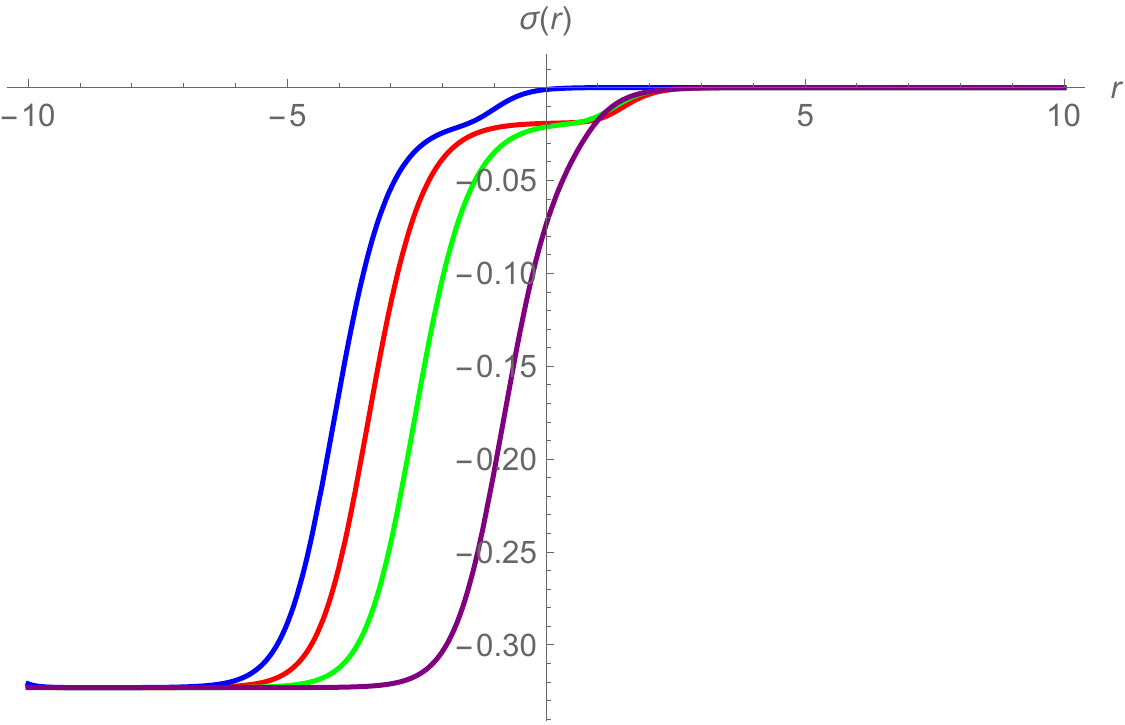}
                 \caption{Solutions for $\sigma(r)$}
         \end{subfigure}
          \begin{subfigure}[b]{0.45\textwidth}
                 \includegraphics[width=\textwidth]{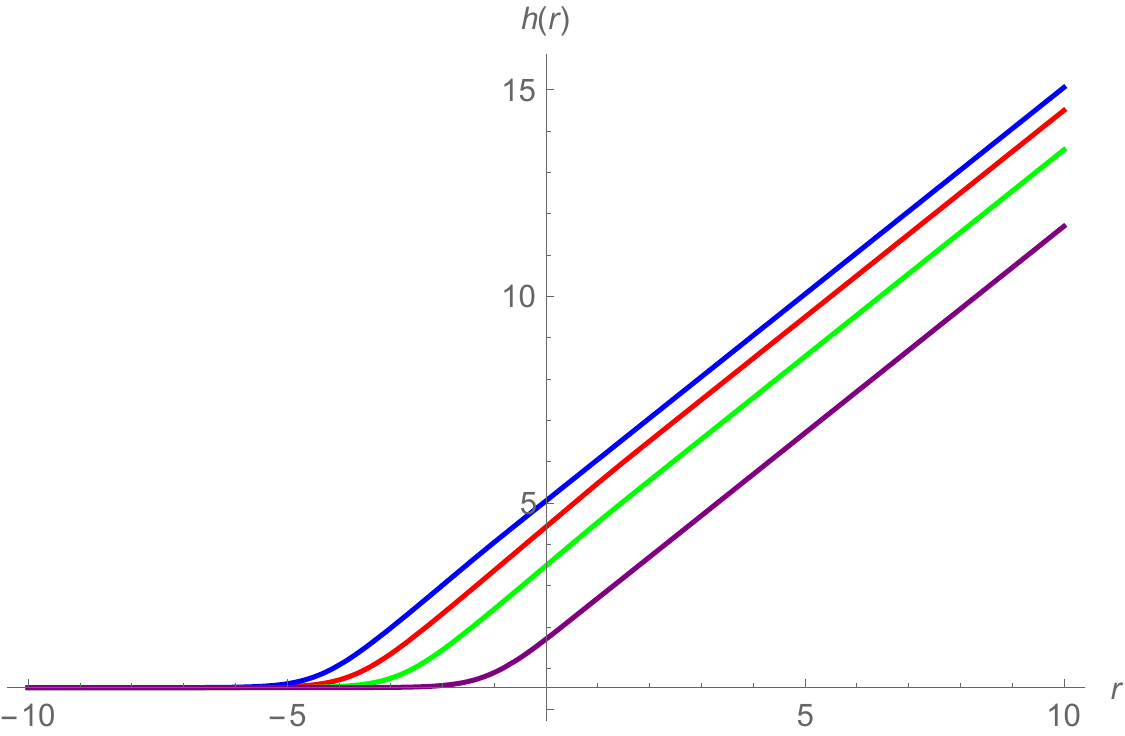}
                 \caption{Solutions for $h(r)$}
         \end{subfigure}\\
          \begin{subfigure}[b]{0.45\textwidth}
                 \includegraphics[width=\textwidth]{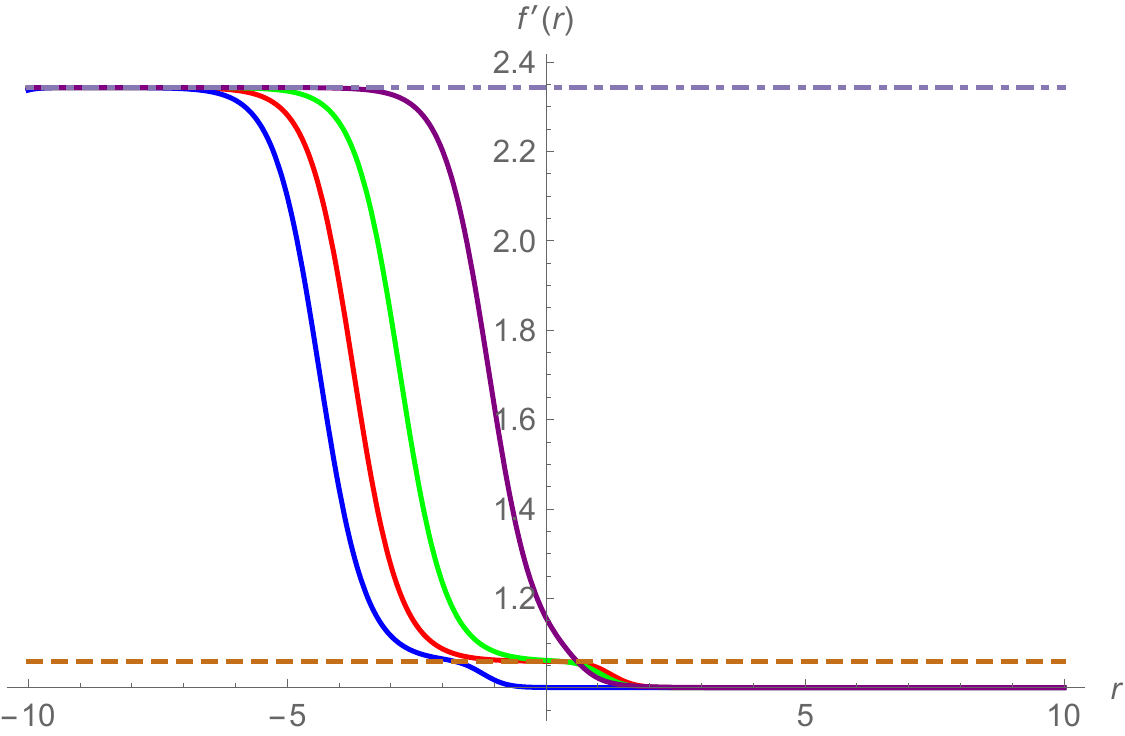}
                 \caption{Solutions for $f'(r)$}
         \end{subfigure}
\caption{Supersymmetric $AdS_6$ black holes interpolating between $AdS_6$ vacua $i$ and $ii$ and the near horizon geometry $AdS_2\times \mc{M}^-_4$ (critical point IV) for $g_2=4$. The purple line represents a solution flowing directly from $AdS_6$ critical point $i$ to $AdS_2\times \mc{M}^-_4$ fixed point IV. The red, green, and blue lines correspond to solutions interpolating between two supersymmetric $AdS_6$ vacua (critical point $i$ and $ii$) and $AdS_2\times \mc{M}^-_4$ geometry IV.}\label{fig4}
 \end{figure}

We first consider the case of $\mc{M}_4$ being a Kahler four-cycle with a $U(2)\sim U(1)\times SU(2)$ holonomy. We will perform the twist along the $U(1)$ part by turning on the $SO(2)_{\textrm{diag}}$ gauge field as in the previous section. The $SO(4,3)/SO(4)\times SO(3)$ coset representative is still given by \eqref{L_SO2diag}. We will choose the metric ansatz to be
\begin{equation}
ds^2=-e^{2f(r)}dt^2+dr^2+e^{2h(r)}ds^2_{\mc{M}_4}
\end{equation}
with the metric on $\mc{M}_4$ given by
\begin{equation}
ds^2_{\mc{M}_4}=\frac{1}{f_\kappa^2(\rho)}\left[d\rho^2+\rho^2f_\kappa(\rho)(\tau_1^2+\tau^2_2)+\rho^2\tau_3^2\right]
\end{equation}
for $f_\kappa(\rho)=1+\kappa \rho^2$. $\tau_i$, $i=1,2,3$, are $SU(2)$ left-invariant one-forms with the normalization
\begin{equation}
d\tau_i=\epsilon_{ijk}\tau_j\wedge \tau_k\, .
\end{equation}
With an obvious choice of vielbein
\begin{eqnarray}
& &e^{\hat{t}}=e^fdt,\qquad e^{\hat{r}}=dr,\qquad e^{\hat{1}}=\frac{e^h\rho}{\sqrt{f_\kappa(\rho)}}\tau_1,\nonumber \\
& &e^{\hat{2}}=\frac{e^h\rho}{\sqrt{f_\kappa(\rho)}}\tau_2,\qquad e^{\hat{3}}=\frac{e^h\rho}{f_\kappa(\rho)}\tau_3,\qquad e^{\hat{4}}=\frac{e^h}{f_\kappa(\rho)}d\rho,
\end{eqnarray}
we can determine non-vanishing components of the spin connection
\begin{eqnarray}
& &{\omega^{\hat{t}}}_{\hat{r}}=f'e^{\hat{t}},\qquad {\omega^{\hat{\alpha}}}_{\hat{r}}=h'e^{\hat{\alpha}},\quad \hat{\alpha}=1,2,3,4,\nonumber \\
& &{\omega^{\hat{1}}}_{\hat{4}}={\omega^{\hat{2}}}_{\hat{3}}=\frac{e^{-h}}{\rho}e^{\hat{1}},\qquad
{\omega^{\hat{2}}}_{\hat{4}}={\omega^{\hat{3}}}_{\hat{1}}=\frac{e^{-h}}{\rho}e^{\hat{2}},\nonumber \\ 
& &{\omega^{\hat{1}}}_{\hat{2}}=\frac{e^{-h}}{\rho}(2\kappa \rho^2+1)e^{\hat{3}},\qquad {\omega^{\hat{4}}}_{\hat{3}}=\frac{e^{-h}}{\rho}(\kappa \rho^2-1)e^{\hat{3}}\, .
\end{eqnarray}
To implement the topological twist along $\mc{M}_4$, we turn on the following gauge fields
\begin{equation}
A^3=3a\kappa\rho e^{-h}e^{\hat{3}}\qquad \textrm{and}\qquad A^6=3b\kappa \rho e^{-h}e^{\hat{3}}
\end{equation}
with $g_2A^6=g_1A^3$ as in the previous section. 
\\
\indent We first consider the $\delta \psi_{\hat{3}A}$ condition
\begin{equation}
0=\frac{1}{2}\frac{e^{-h}}{\rho}(2\kappa\rho^2+1)\gamma_{\hat{1}\hat{2}}\epsilon_A+\frac{1}{2}\frac{e^{-h}}{\rho}(\kappa\rho^2-1)\gamma_{\hat{4}\hat{3}}\epsilon_A-\frac{3i}{2}g_1a\kappa \rho e^{-h}\sigma^3_{AB}\epsilon^B+\ldots
\end{equation}
in which we have used the composite connection given in \eqref{SO2diag_composite}. We now perform the twist by imposing the projectors 
\begin{equation}
\gamma_{\hat{1}\hat{2}}\epsilon_A=-\gamma_{\hat{3}\hat{4}}\epsilon_A=i\sigma^3_{AB}\epsilon^B\label{Proj_Kahler4}
\end{equation}
along with the twist condition
\begin{equation}
g_1a=1\, .
\end{equation}
On the other hand, the conditions $\delta\psi_{\hat{1}A}$ and $\delta\psi_{\hat{2}A}$ give
\begin{equation}
0=\frac{1}{2}\frac{e^{-h}}{\rho}(\gamma_{\hat{1}\hat{4}}+\gamma_{\hat{2}\hat{3}})\epsilon_{A}+\ldots\qquad \textrm{and}\qquad
0=\frac{1}{2}\frac{e^{-h}}{\rho}(\gamma_{\hat{2}\hat{4}}+\gamma_{\hat{3}\hat{1}})\epsilon_{A}+\ldots\, .
\end{equation}
By the projectors \eqref{Proj_Kahler4}, these terms vanish identically, and all the conditions from $\delta\psi_{\hat{\alpha}A}$ with $\hat{\alpha}=1,2,3,4$ lead to the same BPS equation for the warp factor $h(r)$. 
\\
\indent With the gauge field strength tensors
\begin{equation}
F^3=6\kappa ae^{-2h}(e^{\hat{1}}\wedge e^{\hat{2}}-e^{\hat{3}}\wedge e^{\hat{4}})\qquad \textrm{and}\qquad F^6=6\kappa be^{-2h}(e^{\hat{1}}\wedge e^{\hat{2}}-e^{\hat{3}}\wedge e^{\hat{4}}),
\end{equation}
we find the non-vanishing component of the two-form field
\begin{equation}
B_{\hat{t}\hat{r}}=\frac{9}{2}\frac{\kappa^2e^{2\sigma-4h}}{m^2\mc{N}_{00}}(b^2-a^2).
\end{equation}
We also note that the projectors in \eqref{Proj_Kahler4} imply
\begin{equation}
\gamma^{\hat{t}\hat{r}}\epsilon_A=-i\gamma_7\epsilon_A\, .
\end{equation}
Together with the $\gamma^{\hat{r}}$ projector \eqref{gamma_r_proj}, we find the following BPS equations
\begin{eqnarray}
\phi_1'&=&-e^\sigma\textrm{sech}2\phi_3\sinh2\phi_1(g_1\cosh\phi_2-g_2\sinh\phi_2),\\
\phi_2'&=&-e^\sigma\left[g_1\sinh\phi_2+g_2\cosh\phi_2+\cosh2\phi_1\cosh2\phi_3(g_1\sinh\phi_2-g_2\cosh\phi_2)\right]\nonumber \\
& &+6\kappa e^{-\sigma-2h}(b\cosh\phi_2-a\sinh\phi_2),\\
\phi_3'&=&-e^\sigma\cosh2\phi_1\sinh2\phi_3(g_1\cosh\phi_2-g_2\sinh\phi_2),\\
\sigma'&=&-\frac{1}{4}e^\sigma\left[g_1\cosh\phi_2+g_2\sinh\phi_2+\cosh2\phi_1\cosh2\phi_3(g_1\cosh\phi_2-g_2\sinh\phi_2)\right]\nonumber \\
& &+\frac{3}{2}me^{-3\sigma}+\frac{9}{2}\kappa e^{-\sigma-2h}(a\cosh\phi_2-b\sinh\phi_2)+\frac{9\kappa^2}{8m}e^{\sigma-4h}(b^2-a^2),\\
h'&=&\frac{1}{4}e^\sigma\left[g_1\cosh\phi_2+g_2\sinh\phi_2+\cosh2\phi_1\cosh2\phi_3(g_1\cosh\phi_2-g_2\sinh\phi_2)\right]\nonumber \\
& &+\frac{1}{2}me^{-3\sigma}+\frac{9}{2}\kappa e^{-\sigma-2h}(a\cosh\phi_2-b\sinh\phi_2)+\frac{9\kappa^2}{8m}e^{\sigma-4h}(a^2-b^2),\\
f'&=&\frac{1}{4}e^\sigma\left[g_1\cosh\phi_2+g_2\sinh\phi_2+\cosh2\phi_1\cosh2\phi_3(g_1\cosh\phi_2-g_2\sinh\phi_2)\right]\nonumber \\
& &+\frac{1}{2}me^{-3\sigma}-\frac{9}{2}\kappa e^{-\sigma-2h}(a\cosh\phi_2-b\sinh\phi_2)-\frac{27\kappa^2}{8m}e^{\sigma-4h}(a^2-b^2).
\end{eqnarray}
\indent From these equations, we find two $AdS_2\times \mc{M}_4$ critical points:
\begin{itemize}
\item$AdS_2$ critical point III:
\begin{eqnarray}
\phi_1&=&\phi_3=0,\nonumber \\ 
\phi_2&=&\frac{1}{2}\ln\left[\frac{290g_1^2-164g_1g_2-4g_2^2+(g_2-29g_1)\Phi^{\frac{1}{3}}+2\Phi^{\frac{2}{3}}}{9(g_2-g_1)\Phi^{\frac{1}{3}}}\right],\nonumber \\
h&=&\frac{1}{2}\ln\left[\frac{3\kappa e^{-2\sigma}(b\coth\phi_2-a)}{g_1}\right],\nonumber \\
\sigma&=&\frac{1}{4}\ln\left[\frac{2m(b\coth\phi_2-a)}{3g_1(b\sinh\phi_2-a\cosh\phi_2)}\right],\nonumber \\
\frac{1}{\ell}&=&\frac{1}{2}me^{-3\sigma}+2g_1e^\sigma\cosh\phi_2\nonumber \\
& &+\frac{3(b^2-a^2)g_1^2e^{4\sigma}+12bmg_1\textrm{csch}\phi_2(a-b\coth\phi_2)}{8me^{-\sigma}(a-b\coth\phi_2)^2}\label{AdS2_III}
\end{eqnarray}
with
\begin{eqnarray}
\Phi&=&9\sqrt{3}(g_1+g_2)\sqrt{8g_2^4+767g_1^2g_2^2-1000g_1^4}\nonumber \\
& &+44g_2^3+33g_1g_2^2+1689g_1^2g_2-1675g_1^3\, .
\end{eqnarray}
\item $AdS_2$ critical point IV:
\begin{eqnarray}
\phi_2&=&\frac{1}{2}\ln\left[\frac{g_2+g_1}{g_2-g_1}\right],\qquad \kappa=-1,\nonumber \\
h&=&\frac{1}{4}\ln\left[\frac{9}{4}\frac{a^2(g_1^2+8g_2^2)+b^2(8g_1^2+g_2^2)-18abg_1g_2}{mg_1g_2\sqrt{g_2^2-g_1^2}}\right],\nonumber \\
\sigma&=&\frac{1}{4}\ln\left[\frac{m\sqrt{g_2^2-g_1^2}[a^2(g_1^2+8g_2^2)+b^2(8g_1^2+g_2^2)-18abg_1g_2]}{9g_1g_2(bg_1-ag_2)^2}\right],\nonumber \\
\phi_1&=&\frac{1}{2}\ln\left[\frac{2(g_1^2+g_2^2)+\sqrt{2}\sqrt{g_1^4+6g_1^2g_2^2+g_2^4+(g_2^2-g_1^2)^2\cosh4\phi_3}}{2(g_2^2-g_1^2)\cosh2\phi_3}\right],\nonumber \\
\frac{1}{\ell}&=&\frac{2^{\frac{7}{4}}g_1^{\frac{3}{4}}g_2^{\frac{3}{4}}m^{\frac{1}{4}}}{\sqrt{3}(g_2^2-g_1^2)^{\frac{3}{8}}}\label{AdS2_IV}
\end{eqnarray}
with $\phi_3$ being a constant. As in the previous section, for $\phi_3=0$, we find
\begin{equation}
\phi_1=\frac{1}{2}\ln\left[\frac{g_2+g_1}{g_2-g_1}\right].
\end{equation}  
We note again that only $\phi_1$ depends on the value of $\phi_3$.  
\end{itemize}
\indent In both of these critical points, $AdS_2\times \mc{M}_4$ solutions exist only for $\kappa=-1$ leading to black holes with $\mc{M}_{4}^-$ horizons. We now give numerical black hole solutions interpolating between these geometries and supersymmetric $AdS_6$ vacua. With $\phi_1=\phi_3=0$, $m=\frac{1}{2}$, and $\kappa=-1$, we find examples of solutions interpolating between $AdS_6$ vacua $i$ and $AdS_2\times \mc{M}^-_4$ critical point III with different values of $g_2$ as shown in figure \ref{fig3}. Similar to the black hole solutions with $H^2\times H^2$ horizons given in the previous section, we can compute the black hole entropy as
\begin{eqnarray}
S_{\textrm{BH}}&=&\frac{1}{4G_{\textrm{N}}}e^{4h}\textrm{vol}(\mc{M}_4),\nonumber \\
&=&\frac{27\sinh\phi_2(a-b\coth\phi_2)(a\coth\phi_2-b)\textrm{vol}(\mc{M}_4)}{8g_1mG_{\textrm{N}}}
\end{eqnarray}  
with the value of $\phi_2$ at the $AdS_2\times \mc{M}^-_4$ horizon given in \eqref{AdS2_III}.
 \\
 \indent Examples of black hole solutions with the near horizon geometry given by critical point IV are shown in figure \ref{fig4} for $\phi_3=0$ and $g_2=4$. As in the previous section, there are solutions flowing directly from $AdS_6$ critical point $i$ to $AdS_2\times \mc{M}^-_4$ fixed point IV as shown by the purple line. In addition, there exist solutions interpolating between the two supersymmetric $AdS_6$ vacua (critical point $i$ and $ii$) and $AdS_2\times \mc{M}^-_4$ geometry IV given by red, green, and blue lines. In this case, the entropy of the black hole is given by
\begin{equation}
S_{\textrm{BH}}=\frac{9\left[a^2(g_1^2+g_2^2)+b^2(8g_1^2+g_2^2)-18abg_1g_2\right]\textrm{vol}(\mc{M}_4)}{16g_1g_2m\sqrt{g_2^2-g_1^2}G_{\textrm{N}}}\, .
\end{equation}  

\subsection{Black holes with Cayley four-cycle horizon}
\begin{figure}
         \centering
               \begin{subfigure}[b]{0.45\textwidth}
                 \includegraphics[width=\textwidth]{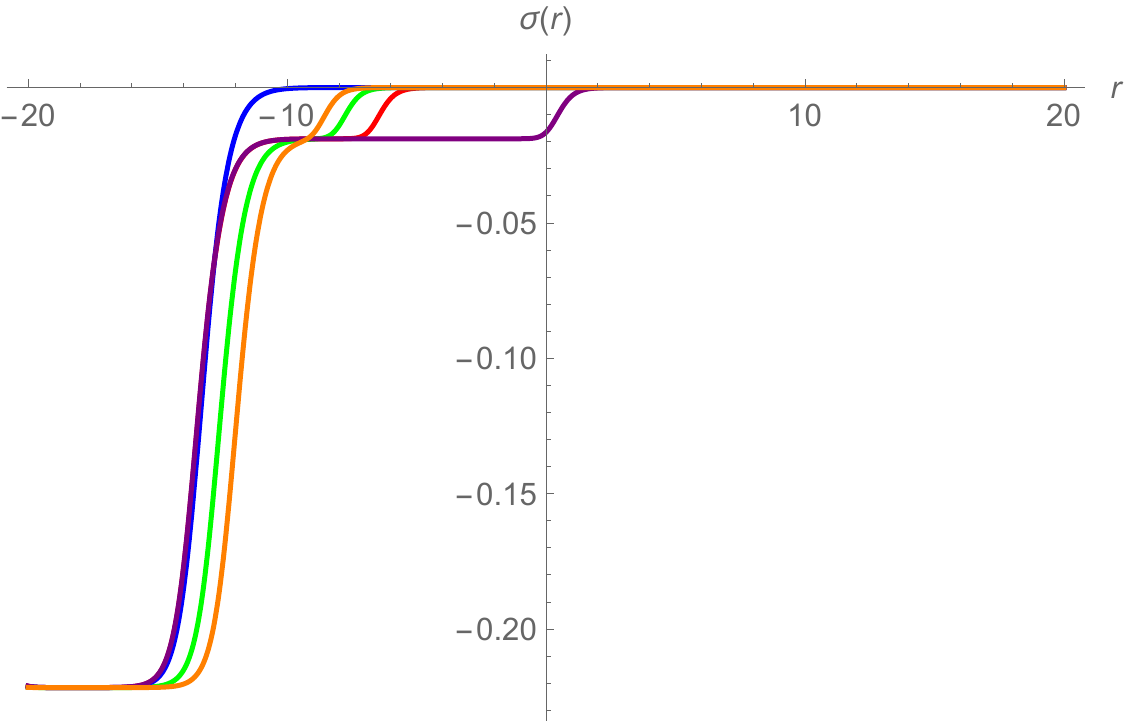}
                 \caption{Solutions for $\sigma(r)$}
         \end{subfigure}
         \begin{subfigure}[b]{0.45\textwidth}
                 \includegraphics[width=\textwidth]{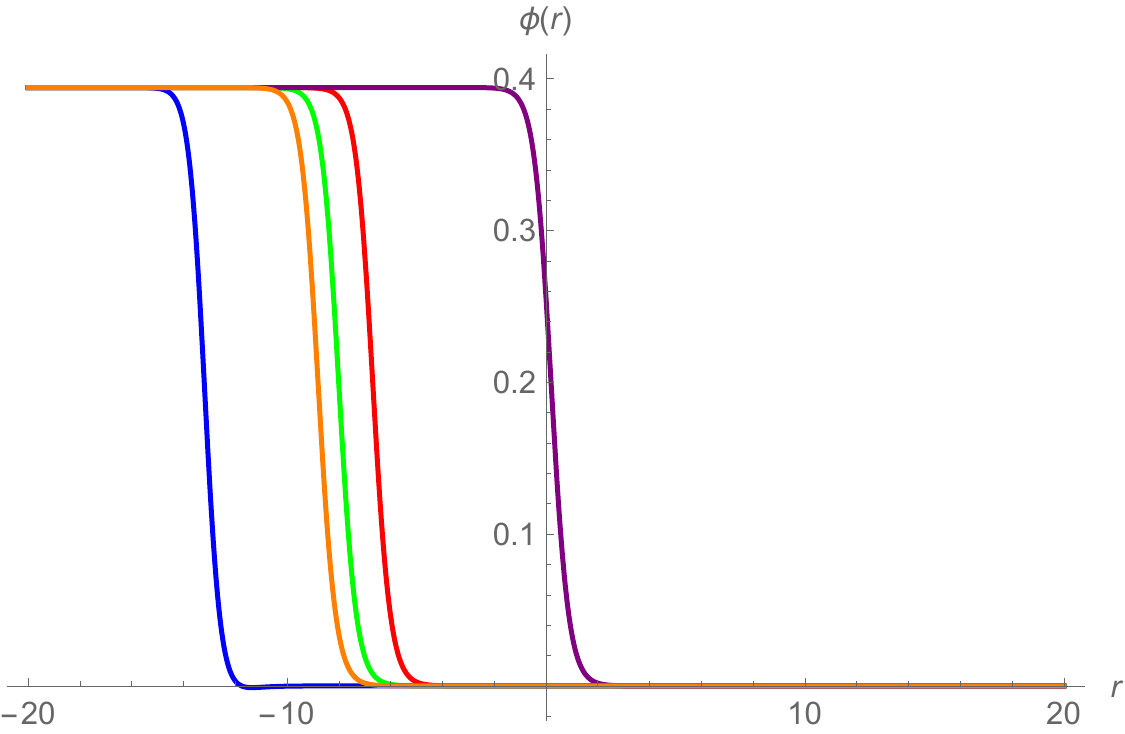}
                 \caption{Solutions for $\phi(r)$}
         \end{subfigure}\\
          \begin{subfigure}[b]{0.45\textwidth}
                 \includegraphics[width=\textwidth]{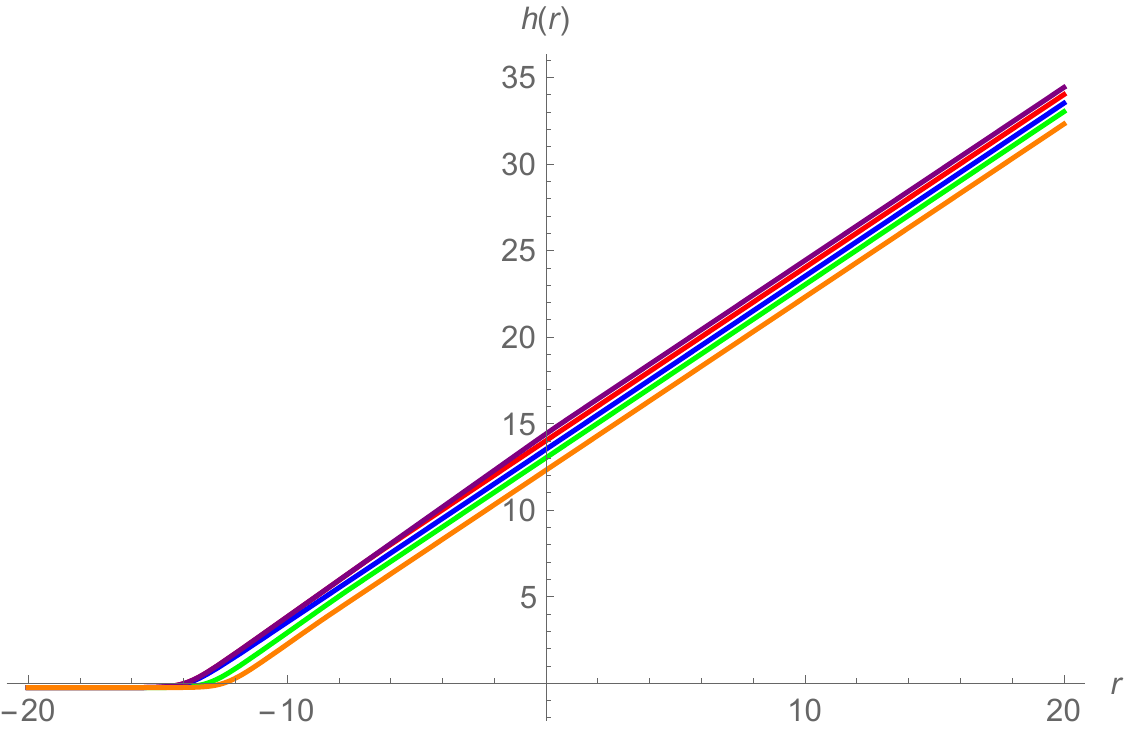}
                 \caption{Solutions for $h(r)$}
         \end{subfigure}
          \begin{subfigure}[b]{0.45\textwidth}
                 \includegraphics[width=\textwidth]{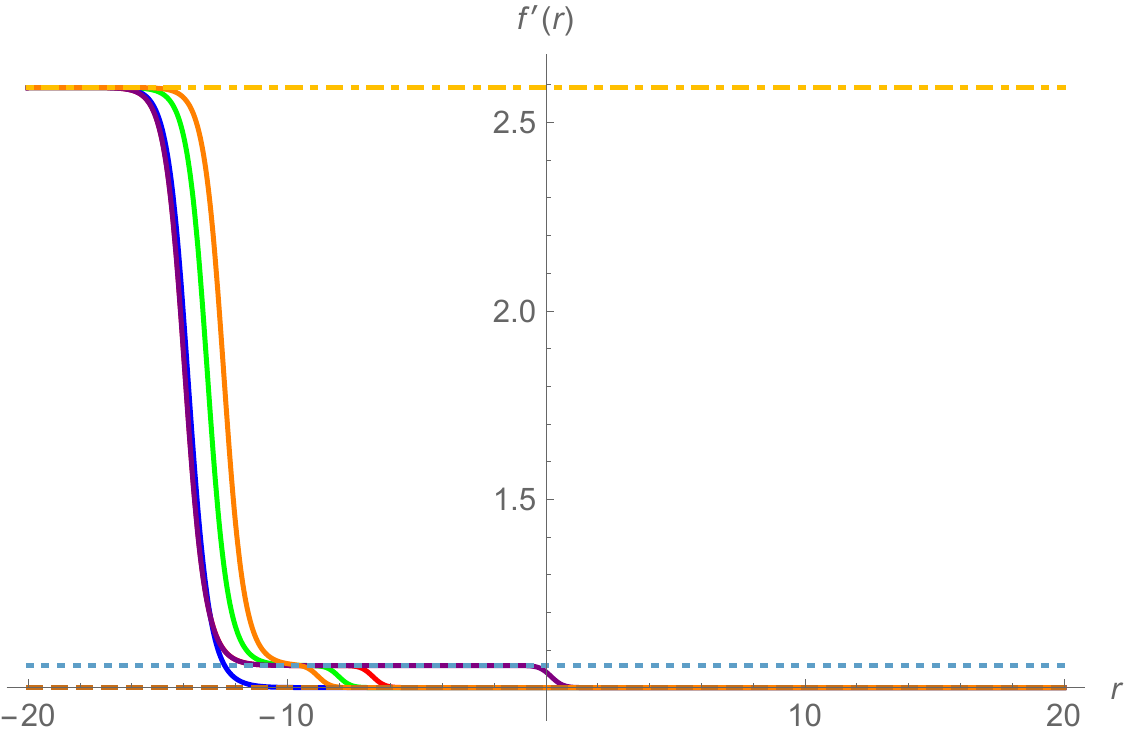}
                 \caption{Solutions for $f'(r)$}
         \end{subfigure}
\caption{Supersymmetric $AdS_6$ black holes interpolating between $AdS_6$ vacua $i$ and $ii$ and the near horizon geometry $AdS_2\times \mc{M}_4^-$ (critical point V) for $m=\frac{1}{2}$ and $g_2=4$. A solution interpolating between $AdS_6$ critical point $i$ and the $AdS_2\times \mc{M}_4^-$ critical point V corresponds to the blue curve while solutions that approach $AdS_6$ critical point $ii$ before flowing to the $AdS_2\times \mc{M}_4^-$ critical point V are shown by orange, green, red, and purple lines.}\label{fig5}
 \end{figure}

In this section, we consider black hole solutions with the horizon $\mc{M}_4$ given by a Cayley four-cycle with the metric ansatz
\begin{equation}
ds^2=-e^{2f(r)}dt^2+dr^2+e^{2h(r)}ds^2_{\mc{M}_4}
\end{equation}
and 
\begin{equation}
ds^2_{\mc{M}_4}=d\rho^2+F_\kappa(\rho)^2(\tau_1^2+\tau_2^2+\tau_3^2)
\end{equation}
in which $\tau_i$ are $SU(2)$ left-invariant one-forms as in the case of Kahler four-cycle and $F_\kappa(\rho)$ defined as in \eqref{F_def}.
\\
\indent In this section, we will perform a twist by turning on $SO(3)_{\textrm{diag}}$ gauge fields and identify this $SO(3)_{\textrm{diag}}$ with the self-dual part $SU(2)_+$ of the $SO(4)\sim SU(2)_+\times SU(2)_-$ isometry of the four-cycle. We note that this has already been considered in \cite{AdS6_BH_Minwoo} in which an $AdS_2\times \mc{M}_4$ solution has been found. However, we do find a new $AdS_2\times \mc{M}_4$ critical point from the resulting BPS equations. Therefore, we will repeat this analysis here with more detail since this might be useful for further study. 
\\
\indent With the vielbein chosen as
\begin{equation}
e^{\hat{t}}=e^fdt,\qquad e^{\hat{r}}=dr,\qquad e^{\hat{i}}=e^hF_\kappa(\rho)\tau_i,\qquad e^{\hat{4}}=e^hd\rho,
\end{equation}
we find non-vanishing components of the spin connection as follows
\begin{eqnarray}
& &{\omega^{\hat{t}}}_{\hat{r}}=f'e^{\hat{r}},\qquad {\omega^{\hat{i}}}_{\hat{r}}=h'e^{\hat{4}},\qquad {\omega^{\hat{i}}}_{\hat{r}}=h'e^{\hat{i}},\qquad \hat{i}=1,2,3,\nonumber \\
& &{\omega^{\hat{i}}}_{\hat{4}}=e^{-h}\frac{F'_\kappa(\rho)}{F_\kappa(\rho)}e^{\hat{i}},\qquad {\omega^{\hat{i}}}_{\hat{j}}=\frac{e^{-h}}{F_\kappa(\rho)}\epsilon_{\hat{i}\hat{j}\hat{k}}e^{\hat{k}}\, .
\end{eqnarray}
Using the coset representative for the $SO(3)_{\textrm{diag}}$ singlet scalar given in \eqref{SO3diag_L}, we find the composite connection
\begin{equation}
Q_{AB}=-\frac{i}{2}g_1A^r\sigma^r_{AB}
\end{equation}
in which we have used the relation $g_2A^I=g_1\delta^I_rA^r$ for $r,I=1,2,3$. In order to cancel the internal spin connection on $\mc{M}_4$, we choose the gauge fields to be
\begin{equation}
A^r=a(F'_\kappa(\rho)+1)\delta^r_i\tau_i\qquad \textrm{and}\qquad A^I=b(F'_\kappa(\rho)+1)\delta^I_i\tau_i
\end{equation}
with $g_2b=g_1a$. We also note that it is possible to begin with three different magnetic charges $a_r$, but the twist condition will subsequently require all the charges to be equal, see also \cite{AdS6_BH_Minwoo}.
\\
\indent We now consider the supersymmetry conditions $\delta \psi_{\hat{i}A}$ which give
\begin{equation}
0=\frac{1}{2}\frac{e^{-h}}{F_\kappa(\rho)}\left(F'_\kappa(\rho)\gamma_{\hat{i}\hat{4}}\epsilon_A+\frac{1}{2}\epsilon_{\hat{i}\hat{j}\hat{k}}\gamma_{\hat{j}\hat{k}}\epsilon_A\right)-\frac{i}{2}g_1A^r_{\hat{i}}\sigma^r_{AB}\epsilon^B+\ldots \, .
\end{equation}
We then impose the following projectors 
\begin{equation}
\gamma_{\hat{i}\hat{4}}\epsilon_A=\frac{1}{2}\epsilon_{\hat{i}\hat{j}\hat{k}}\gamma_{\hat{j}\hat{k}}\epsilon_A=-i\delta^r_i\sigma^r_{AB}\epsilon^B
\end{equation}
and the twist condition
\begin{equation}
g_1a=-1\, .
\end{equation}
With all these, $\delta \psi_{\hat{i}A}$ conditions reduce to the same BPS equation for $h(r)$ obtained from $\delta\psi_{\hat{4}A}$. It should be noted that there are only three independent projectors leading to $\frac{1}{8}$-BPS $AdS_2\times \mc{M}_4$ solutions preserving two supercharges. The full black hole solutions with running scalars will further break supersymmetry to $\frac{1}{16}$-BPS due to the additional $\gamma^{\hat{r}}$ projector \eqref{gamma_r_proj}.
\\
\indent The gauge field strength tensors in this case are given by
\begin{equation}
F^r=\kappa a\delta^r_ie^{-2h}\left(e^{\hat{i}}\wedge e^{\hat{4}}+\frac{1}{2}\epsilon_{\hat{i}\hat{j}\hat{k}}e^{\hat{j}}\wedge e^{\hat{k}}\right)
\end{equation} 
together with $F^I=\frac{g_1}{g_2}\delta^I_rF^r$. The two-form field takes the form
\begin{equation}
B_{\hat{t}\hat{r}}=\frac{3}{8}\frac{\kappa^2e^{2\sigma-4h}}{m^2\mc{N}_{00}}(a^2-b^2).
\end{equation}
With all these and a useful (not independent) relation 
\begin{equation}
\gamma^{\hat{t}\hat{r}}\epsilon_A=i\gamma_7\epsilon_A,
\end{equation}
we obtain the BPS equations
\begin{eqnarray}
\phi'&=&-e^\sigma\sinh2\phi(g_1\cosh\phi-g_2\sinh\phi)+\kappa e^{-\sigma-2h}(a\sinh\phi-b\cosh\phi),\\
\sigma'&=&\frac{3}{2}me^{-3\sigma}-\frac{1}{2}e^\sigma(g_1\cosh^3\phi-g_2\sinh^3\phi)+\frac{3}{4}\kappa e^{-\sigma-2h}(b\sinh\phi-a\cosh\phi)\nonumber \\
& &+\frac{3}{32m}\kappa^2e^{\sigma-4h}(a^2-b^2),\\
h'&=&\frac{1}{2}e^\sigma(g_1\cosh^3\phi-g_2\sinh^3\phi)+\frac{1}{2}me^{-3\sigma}\nonumber \\
& &+\frac{3}{4}\kappa e^{-\sigma-2h}(b\sinh\phi-a\cosh\phi)+\frac{3}{32m}\kappa^2e^{\sigma-4h}(b^2-a^2),\\
f'&=&\frac{1}{2}e^\sigma(g_1\cosh^3\phi-g_2\sinh^3\phi)+\frac{1}{2}me^{-3\sigma}\nonumber \\
& &-\frac{3}{4}\kappa e^{-\sigma-2h}(b\sinh\phi-a\cosh\phi)-\frac{9}{32m}\kappa^2e^{\sigma-4h}(b^2-a^2).
\end{eqnarray}
There are two $AdS_2\times \mc{M}_4$ critical points to these equations. The first one is given by
\begin{eqnarray}
h&=&\frac{1}{2}\ln\left[\frac{3\kappa e^{2\sigma}(a\cosh\phi-b\sinh\phi)}{4m}\right],\nonumber \\
\sigma&=&\frac{1}{4}\ln\left[\frac{4m(b\cosh\phi-a\sinh\phi)}{\sinh2\phi(b\sinh\phi-a\cosh\phi)(g_1\cosh\phi-g_2\sinh\phi)}\right],\nonumber \\
\phi&=&\frac{1}{2}\ln\left[\frac{2\Phi^{\frac{2}{3}}+(13g_2-7g_1)\Phi^{\frac{1}{3}}+2(7g_1^2-50g_1g_2+52g_2^2)}{3(g_1-g_2)\Phi^{\frac{1}{3}}}\right],\nonumber \\
\frac{1}{\ell}&=&\frac{3}{2}me^{-3\sigma}+\frac{1}{2}e^\sigma(g_1\cosh^3\phi-g_2\sinh^3\phi)+\frac{m(a^2-b^2)e^{-3\sigma}}{2(a\cosh\phi-b\sinh\phi)^2}\quad \label{AdS2_CY_Minwoo}
\end{eqnarray}
with 
\begin{equation}
\Phi=3\sqrt{3}(g_2-g_1)\sqrt{131g_1^2g_2^2-192g_2^4-2g_1^4}-17g_1^3+213g_2^2g_2-537g_1g_2^2+368g_2^3\, .
\end{equation}
The solution only exists for $\kappa=-1$. This is the $AdS_2\times \mc{M}^-_4$ solution found in \cite{AdS6_BH_Minwoo} in which the numerical black hole solution has also been given. Therefore, we will not discuss this solution any further but only give, for completeness, the corresponding black hole entropy
\begin{equation}
S_{\textrm{BH}}=\frac{9e^{4\sigma}(a\cosh\phi-b\sinh\phi)^2\textrm{vol}(\mc{M}_4^-)}{64m^2G_{\textrm{N}}}
\end{equation}
with $\phi$ given in \eqref{AdS2_CY_Minwoo}.
\\
\indent There is another $AdS_2\times \mc{M}_4$ solution that has not been given in \cite{AdS6_BH_Minwoo}. This critical point takes the form
\begin{itemize}
\item $AdS_2$ critical point V:
\begin{eqnarray}
\phi&=&\frac{1}{2}\ln\left[\frac{g_2+g_1}{g_2-g_1}\right],\quad \sigma=\frac{1}{4}\ln\left[\frac{4m\sqrt{g_2^2-g_1^2}}{3g_1g_2}\right],\nonumber \\
h&=&\frac{1}{2}\ln\left[\frac{\sqrt{3}\kappa m g_1a(g_2^2-g_1^2)^{\frac{3}{4}}}{2(mg_1g_2)^{\frac{3}{2}}}\right],\qquad \frac{1}{\ell}=\frac{2\sqrt{2}m}{3^{\frac{1}{4}}(g_2^2-g_1^2)^{\frac{3}{8}}}\left(\frac{g_1g_2}{m}\right)^{\frac{3}{4}}\, .\qquad \label{AdS2_V}
\end{eqnarray}
\end{itemize}
As in all the other cases, the solution only exists for $\kappa=-1$. Examples of numerical solutions for $m=\frac{1}{2}$ and $g_2=4$ are given in figure \ref{fig5}. Similar to the previous cases, there are solutions interpolating between $AdS_6$ critical point $i$ and the $AdS_2\times \mc{M}_4^-$ critical point V represented by the blue curve. The orange, green, red, and purple lines correspond to solutions that flow from $AdS_6$ critical point $i$ and approach $AdS_6$ critical point $ii$ before flowing to the $AdS_2\times \mc{M}_4^-$ critical point V. 
\\
\indent The entropy of the black hole is given by
\begin{eqnarray}
S_{\textrm{BH}}=\frac{3a^2(g_2^2-g_1^2)^{\frac{3}{2}}\textrm{vol}(\mc{M}_4)}{16mg_1g_2^3G_{\textrm{N}}}\, .
\end{eqnarray}
\\
\indent We end this section by pointing out that we have also considered topological twists by $SO(2)_R$ and $SO(3)_R$ in the cases of $\mc{M}_4$ being Kahler four-cycle and Cayley four-cycle, respectively. It turns out that there are no new $AdS_2\times \mc{M}_4$ solutions apart from those given in \cite{AdS6_BH_Minwoo} with all scalars in the vector multiplets vanishing.   
\section{Conclusions}\label{conclusion}
We have constructed new supersymmetric $AdS_6$ black hole solutions from $F(4)$ gauged supergravity in six dimensions coupled to three vector multiplets with $SO(3)\times SO(3)$ gauge group. By considering a truncation to $SO(2)_{\textrm{diag}}$ singlet sector and performing a topological twist by the $SO(2)_{\textrm{diag}}$ gauge field, we find a number of black hole solutions with $AdS_2\times H^2\times H^2$ and $AdS_2\times \mc{M}^-_4$ near horizon geometries with $\mc{M}_4^-$ being a negatively curved Kahler four-cycle. On the other hand, by performing a twist by $SO(3)_{\textrm{diag}}$ gauge fields, we have found a new black hole solution with $AdS_2\times \mc{M}_4^-$ near horizon geometry for $\mc{M}^-_4$ being a negatively curved Cayley four-cycle. All the solutions identified in this paper are collectively shown in table \ref{table1}.

\begin{table}[!htbp]
\renewcommand{\arraystretch}{1.5}
\centering
\begin{tabular}{|c|c|c|}
\hline
Solution & Near horizon geometry & Numerical black hole solution\\
\hline
Critical point I & $AdS_2\times H^2\times H^2$ \eqref{AdS2_Sigma_Sigma1} &  Figure \ref{fig1}\\ \hline
Critical point II & $AdS_2\times H^2\times H^2$ \eqref{AdS2_Sigma_Sigma2} &  Figure \ref{fig2}\\ \hline
Critical point III & $AdS_2\times \mc{M}^-_{\textrm{K}4}$ \eqref{AdS2_III} &  Figure \ref{fig3}\\ \hline
Critical point IV & $AdS_2\times \mc{M}^-_{\textrm{K}4}$ \eqref{AdS2_IV} &  Figure \ref{fig4}\\ \hline
Critical point V & $AdS_2\times \mc{M}^-_{\textrm{C}4}$ \eqref{AdS2_V} &  Figure \ref{fig5}\\ \hline
\end{tabular}
\caption{Near horizon $AdS_2\times \mc{M}_4$ geometries and numerical solutions for supersymmetric $AdS_6$ black holes found in this paper. The notations $\mc{M}^-_{\textrm{K}4}$ and $\mc{M}^-_{\textrm{C}4}$ correspond to $\mc{M}_4$ being Kahler 4-cycle and Cayley four-cycle, respectively.}\label{table1}
\end{table}
\indent We have also given examples of numerical black hole solutions interpolating between these near horizon geometries and the two supersymmetric $AdS_6$ vacua. Unlike the previously found solutions in \cite{AdS6_BH_Zaffaroni} and \cite{AdS6_BH_Minwoo}, some of the solutions given in this paper interpolate between the near horizon geometries and both of the supersymmetric $AdS_6$ vacua. According to the AdS/CFT correspondence, these solutions should describe RG flows across dimensions to superconformal quantum mechanics arising from twisted compactifications on $\mc{M}_4$ of five-dimensional SCFTs dual to the $AdS_6$ vacua. We hope the new black hole solutions given here would be useful in the study of attractor mechanism in six dimensions and provide a holographic description of twisted compactifications of the dual five-dimensional SCFTs on four-manifolds.
\\
\indent It would be interesting to compute the black hole entropy for the solutions given here from the topologically twisted indices of the dual SCFTs in five dimensions using the results of \cite{5D_twist_index1,5D_twist_index2} possibly with some extension and modification. It is of particular interest to find possible embedding of the solutions found here and in \cite{AdS6_BH_Zaffaroni,AdS6_BH_Minwoo} in ten or eleven dimensions. Along this direction, the results of \cite{Malek_AdS6_1} and \cite{Malek_AdS6_2} might be useful. Finding $AdS_6$ black hole solutions with the horizons being four-dimensional orbifolds as in the recent results \cite{spindle_BH1,spindle_BH2,4_orbifold_BH1,4_orbifold_BH2,4_orbifold_BH3} is also worth considering. We leave these issues for future works.             
\vspace{0.5cm}\\
{\large{\textbf{Acknowledgement}}} \\
This work is funded by National Research Council of Thailand (NRCT) and Chulalongkorn University under grant N42A650263. The author would also like to thank C. A. for mental supports during hard times in his life.
\appendix
\section{Bosonic field equations of matter-coupled $F(4)$ gauged supergravity}
In this appendix, we collect all the bosonic field equations obtained from the Lagrangian given in \eqref{Lar}. These are given by 
\begin{eqnarray}
D_\mu D^\mu \sigma-\frac{1}{4}e^{-2\sigma}\mc{N}_{\Lambda\Sigma}\widehat{F}_{\mu\nu}^\Lambda \widehat{F}^{\Sigma \mu\nu}+\frac{3}{16}e^{4\sigma}H_{\mu\nu\rho}H^{\mu\nu\rho}\qquad \qquad & &\nonumber \\ -6me^{-6\sigma}\mc{N}_{00}+4me^{-2\sigma}\left(\frac{1}{3}AL_{00}-L_{0i}B^i\right)\qquad & &\nonumber \\
-e^{2\sigma}\left[\frac{1}{18}A^2+\frac{1}{2}B^iB_i+\frac{1}{2}(C_{It}C^{It}+4D_{It}D^{It})\right]&=&0,\\
\frac{1}{2}D_\mu P^{\mu I0}-e^{-2\sigma}L_{0\Lambda}L_{I\Sigma}\widehat{F}^\Lambda_{\mu\nu}\widehat{F}^{\Sigma\mu\nu}-2me^{-2\sigma}L_{0r}C_{Ir}\qquad & &\nonumber \\
+e^{2\sigma}\left(B_iC_{iI}+K_{tIJ}D^{Jt}\right)&=&0,\\
\frac{1}{2}D_\mu P^{\mu Ir}+2me^{-2\sigma}\left(L_{00}C_{Ir}
-2\epsilon_{rst}L_{0s}D_{It}\right)\qquad \qquad\qquad& &\nonumber \\
-e^{-2\sigma}L_{r\Lambda}L_{I\Sigma}\widehat{F}^\Lambda_{\mu\nu}\widehat{F}^{\Sigma\mu\nu}+e^{2\sigma}\left[\frac{1}{3}AC_{Ir}+\epsilon_{rst}(B^sD_{It}+C_{Js}K_{tIJ})\right]&=&0,\\
D_\nu\left(e^{-2\sigma}\mc{N}_{\Lambda\Sigma}\widehat{F}^{\Sigma \nu\mu}\right)+2P^\mu_{I\alpha}{(L^{-1})^I}_\Sigma {{f_\Lambda}^\Sigma}_\Pi{L^\Pi}_\alpha\qquad \qquad & &\nonumber \\
-\frac{1}{8}\epsilon^{\mu\rho\sigma\nu\lambda\tau}H_{\rho\sigma\nu}\left(mB_{\lambda\tau}\delta^{\Lambda 0}+\eta_{\Lambda\Sigma}F^\Sigma_{\lambda\tau}\right)&=&0,\\
\frac{3}{8}D_\rho\left(e^{4\sigma}H^{\mu\nu\rho}\right)-m^2e^{-2\sigma}\mc{N}_{00}B^{\mu\nu}+me^{-2\sigma}\mc{N}_{0\Sigma}F^{\Sigma \mu\nu}\qquad\qquad & &\nonumber \\
-\frac{1}{16}\epsilon^{\mu\nu\rho\sigma\lambda\tau}\left(\eta_{\Lambda\Sigma}F^\Lambda_{\rho\sigma}F^\Sigma_{\lambda\tau}
-2mF^0_{\lambda\tau}B_{\rho\sigma}+m^2B_{\rho\sigma}B_{\lambda\tau}\right)&=&0,\label{B_eq}\\
R_{\mu\nu}-\frac{1}{2}g_{\mu\nu}R-e^{-2\sigma}\mc{N}_{\Lambda\Sigma}\widehat{F}^\Lambda_{\mu\rho}\widehat{F}^{\Sigma\phantom{\nu}\rho}_{\phantom{\Sigma}\nu}-4\pd_\mu \sigma\pd_\nu \sigma \qquad \qquad\qquad\qquad & &\nonumber \\
-P^{I\alpha}_\mu P_{\nu I\alpha}-\frac{9}{16}e^{4\sigma}H_{\mu\rho\sigma}{H_\nu}^{\rho\sigma}
+g_{\mu\nu}\left[\frac{1}{4}e^{-2\sigma}\mc{N}_{\Lambda\Sigma}\widehat{F}^\Lambda_{\rho\sigma}\widehat{F}^{\Sigma\rho\sigma}\right.  \qquad & & \nonumber \\
\left. +\frac{3}{32}e^{4\sigma}H_{\rho\sigma\lambda}H^{\rho\sigma\lambda}+2\pd_\rho\sigma\pd^\rho\sigma+\frac{1}{2}P^{I\alpha}_\rho P^\rho_{I\alpha}+2V\right]&=&0\qquad 
\end{eqnarray}
with 
\begin{equation}
K_{rIJ}=g_1\epsilon_{lmn}{L^l}_r{(L^{-1})_I}^m{L^n}_J+g_2C_{LMK}{L^L}_r{(L^{-1})_I}^M{L^K}_J\, .
\end{equation}
\\
\textbf{Data Availability Statement:} No data associated in the manuscript

\end{document}